\documentclass[twocolumn]{aastex631}

\usepackage{amsmath}
\usepackage{longtable}
\usepackage{array}

\shorttitle{CAMELS Feedback Implementation}
\shortauthors{Medlock \& Neufeld et al.}

\graphicspath{{./}{figures/}}

\begin{document}

\title{Quantifying Baryonic Feedback on Warm-Hot Circumgalactic Medium in CAMELS Simulations}

\author{Isabel Medlock}
\affiliation{Department of Astronomy, Yale University, New Haven, CT 06520, USA}

\author{Chloe Neufeld}
\affiliation{Department of Astronomy, Yale University, New Haven, CT 06520, USA}

\author{Daisuke Nagai}
\affiliation{Department of Astronomy, Yale University, New Haven, CT 06520, USA}
\affiliation{Department of Physics, Yale University, New Haven, CT 06520, USA}

\author{Daniel Anglés-Alcázar}
\affiliation{Department of Physics, University of Connecticut, 196 Auditorium Road, U-3046, Storrs, CT, 06269, USA}

\author{Shy Genel}
\affiliation{Center for Computational Astrophysics, Flatiron Institute, 162 5th Avenue, New York, NY, 10010, USA}
\affiliation{Columbia Astrophysics Laboratory, Columbia University, 550 West 120th Street, New York, NY, 10027, US}

\author{Benjamin D. Oppenheimer}
\affiliation{CASA, Department of Astrophysical and Planetary Sciences, University of Colorado, Boulder, CO 80309, USA}

\author{Xavier Sims}
\affiliation{Department of Physics, University of Connecticut, 196 Auditorium Road, U-3046, Storrs, CT, 06269, USA}

\author{Priyanka Singh}
\affiliation{Department of Astronomy, Astrophysics \& Space Engineering, Indian Institute of Technology Indore, India}
\affiliation{Department of Physics, Yale University, New Haven, CT 06520, USA}

\author{Francisco Villaescusa-Navarro}
\affiliation{Center for Computational Astrophysics, Flatiron Institute, 162 5th Avenue, New York, NY, 10010, USA}
\affiliation{Department of Astrophysical Sciences, Princeton University, Peyton Hall, Princeton, NJ, 08544, USA}

\begin{abstract}

The baryonic physics shaping galaxy formation and evolution are complex, spanning a vast range of scales and making them challenging to model. Cosmological simulations rely on subgrid models that produce significantly different predictions. Understanding how models of stellar and active galactic nuclei (AGN) feedback affect baryon behavior across different halo masses and redshifts is essential. Using the SIMBA and IllustrisTNG suites from the Cosmology and Astrophysics with MachinE Learning Simulations (CAMELS) project, we explore the effect of parameters governing the subgrid implementation of stellar and AGN feedback. We find that while IllustrisTNG shows higher cumulative feedback energy across all halos, SIMBA demonstrates a greater spread of baryons, quantified by the closure radius and circumgalactic medium (CGM) gas fraction. This suggests that feedback in SIMBA couples more effectively to baryons and drives them more efficiently within the host halo. There is evidence that different feedback modes are highly interrelated in these subgrid models. Parameters controlling stellar feedback efficiency significantly impact AGN feedback, as seen in the suppression of black hole mass growth and delayed activation of AGN feedback to higher mass halos with increasing stellar feedback efficiency in both simulations. Additionally, AGN feedback efficiency parameters affect the CGM gas fraction at low halo masses in SIMBA, hinting at complex, non-linear interactions between AGN and SNe feedback modes. Overall, we demonstrate that stellar and AGN feedback are intimately interwoven, especially at low redshift, due to subgrid implementation, resulting in halo property effects that might initially seem counterintuitive.

\end{abstract}


\section{Introduction} \label{sec:intro}

Galaxy evolution is governed by complex physical processes across various spatial and dynamic scales. In particular, star formation and black hole growth along with their associated feedback mechanisms are important drivers of this evolution. Recent advances in computational methods and observational capabilities have provided new insight into these feedback processes. As black hole feedback is related to the co-evolution between black holes and host galaxies \citep[e.g.,][]{croton_agnfeedback_2006,hopkins_agn_2008}, the effects of these physical processes are imprinted in the characteristics of the host halo \citep{cattaneo_role_2009,harrison_agn_2018}. 

However, the exact physical impact on host halos is not yet fully understood due to the complex physics involved and the scales at which these processes exist (see \citealp{somerville_physical_2015}, \citealp{tumlinson_circumgalactic_2017}, \citealp{Vogelsberger_2020}, and \citealp{Crain_2023} for reviews on the topic). To address these challenges, cosmological simulations often incorporate subgrid physics, particularly in modeling star formation and active galactic nuclei (AGN) feedback. Although these models are tuned to match observables such as the stellar halo mass relation, they predict significantly different behavior, in particular in the distribution of gas around galaxies \citep[e.g.,][]{Chisari_2018, Chisari_2019, vanDaalen_2020, Oppenheimer_2021, Gebhardt_2024}. Thus, understanding the implementation of these subgrid prescriptions is crucial for meaningful comparisons with other models or observations, as well as for extracting the full physical significance of simulation results. 

Several complex and interacting processes make up these feedback mechanisms, beginning from the growth of stellar and black hole components in a galaxy to the outflows and energy injection that subsequently occur from these processes. Black holes grow from nuclear gas inflows driven by gravitational instabilities and galaxy mergers \citealp{Alexander_2012}. Star formation occurs as gas collapses, becomes self-gravitating, cools, forms a giant molecular cloud that fragments into dense cloud cores, and eventually reaches densities high enough to ignite nuclear fusion \citep{kennicutt_sfr_1998}. Feedback from massive stars and AGN can then be divided into two types: preventative and ejective. Preventative feedback refers to processes that prevent gas accretion onto the interstellar medium (ISM), thus slowing star formation, and ejective feedback refers to those that remove gas from the ISM after it is accreted. 

Stellar feedback is thought to affect the efficiency at which gas is converted into stars, which indeed has been observed to be lower than expected (e.g., \citealp{krumholz_sf_2015} find efficiencies of $\sim 1$\% per free-fall time). The baryon fraction within the halos is also smaller than expected. For example, \cite{behroozi_stellarmass_2010} and \cite{moster_stellarmass_2010} find baryon fractions less than 20\% of the universal value. Star formation processes that could lead to such inefficiencies are supernova (SNe) explosions, which can heat and expel gas \citep{white_rees_1978,dekel_silk_1986,hopkins_starfeedback_2021}, preventing gas from cooling (and thus stars from forming) and reducing the baryon fraction in galaxies.

Moreover, most massive galaxies contain a supermassive black hole (SMBH) \citep{kormendy_ho_2013}, and the large amounts of feedback energy released during BH growth are thought to have a strong effect on galaxy evolution \citep{silk_rees_1998,fabian_2012,heckman_2014}. These outflows have been confirmed observationally \citep[e.g.,][]{rupke_2011,maiolino_2012,harrison_2014,cicone_2015,cicone_2016} and investigated through simulations \citep[e.g.,][]{springel_modelling_2005,hopkins_2005,booth_2009,johansson_2009,dubois_2013, Mercedes-Feliz_2023}, with effects ranging from high-velocity winds that expel gas to radio jets that form hot bubbles and heat the surrounding gas.

Acting as the interface of these baryonic flows is the complex and multiphase circumgalactic medium (CGM), the diffuse reservoir of gas beyond the ISM extending out to (and possibly beyond) the halo's virial radius. The boundaries of the CGM are nebulous and difficult to define. Recently, \cite{aryomlou2023} introduced the concept of the closure radius to define the radial extent of the impact of feedback. Additionally, the fraction of mass that is in the CGM, defined as $f_{\rm CGM}$, is determined by these feedback processes \citep[e.g.,][]{best_agn_2007,voit_agn_2017,davies_gas_2019,davies_quenching_2020}. However, due to its diffuse nature, the CGM is challenging to observe and is mainly limited to absorption line studies and thus quasar-galaxy pairs \citep[e.g.,][]{matejek2012survey, lehner2015evidence, rubin2015dissecting, bowen2016structure, turner2017monthly}, or "down the barrel" observations \citep[e.g.,][]{martin2005mapping, steidel2010structure, bordoloi2011radial, kornei2012properties, rubin2014evidence, heckman_2014, henry2015lyalpha}, both of which provide limited spatial information. Alternatively, we can use simulations to explore the CGM in the context of feedback processes to inform future observations.

Multiple simulations have been used to explore feedback effects in relation to galaxy quenching and the CGM. Using the Feedback in Realistic Environments (FIRE; \citealp{hopkins_fire_2014,hopkins_fire_2018,hopkins_fire-3_2023}) cosmological zoom simulations and exploring several modeling choices for SMBH accretion and AGN feedback, \cite{wellons_exploring_2023} find that the regulation of black hole growth and star formation in galaxies is closely related, with correlations between stellar mass growth, black hole growth, accretion efficiency, and feedback efficiencies. \cite{davies_quenching_2020} use IllustrisTNG \citep{pillepich_illustris_sim_2018,pillepich_illustris_stellarmass_2018,nelson_illustris_galaxycolor_2018,springel_illustris_clustering_2018} and EAGLE \citep{crain_eagle_2015,Schaye_2015} simulations, finding a negative correlation between feedback energy and halo mass, as well as a correlation between high feedback energy and low $f_{\rm CGM}$. Furthermore, \cite{oppenheimer_agn_2020} find that $f_{\rm CGM}$ decreases due to outflows from AGN feedback. 

Although these results are illuminating, it is difficult to directly compare the results and conclusions due to not only the difference in subgrid implementation of the simulations but also the setup including the initial conditions, resolution, and volume. The Cosmology and Astrophysics with MachinE Learning Simulations (CAMELS) provides a set of simulations that normalizes all of these specifics while varying subgrid models and cosmological and astrophysical feedback parameters \citep{camels_presentation}. This work explores stellar and black hole feedback within two suites of cosmological simulations from the CAMELS project: SIMBA \citep{dave_simba_2019} and IllustrisTNG. We analyze 96 different simulations with different subgrid models and parameter values. We quantify the cumulative feedback energy from redshift $z\sim3$ to the present epoch from the different stellar and AGN feedback modes. We analyze correlations between feedback energy, halo mass, $f_{\rm CGM}$, and closure radius to gain a better understanding of the underlying processes that drive galaxy evolution, specifically analyzing the effects of different feedback implementations on the baryons within a host halo.

The paper is structured as follows. We review the relevant background and key quantities in \S\ref{sec:back}. In \S\ref{sec:data}, we provide an overview of the CAMELS project (\S\ref{sec:camels}) as well as in-depth reviews of the feedback implementation in the CAMELS versions of SIMBA (\S\ref{sec:simba}) and IllustrisTNG (\S\ref{sec:tng}). We present our findings in \S\ref{sec:results} and in \S\ref{sec:disc} discuss the caveats and next steps. Finally, we summarize our key conclusions and takeaways in \S\ref{sec:conclusion}.

\section{Background} \label{sec:back}

One straightforward approach to understanding feedback processes is to directly calculate the cumulative feedback energy of each mode as implemented in a given simulation. We can then compare the relative contribution of each mode as a function of halo mass and redshift as well as investigate correlations with key CGM properties. \cite{davies_quenching_2020} completed this analysis with the original EAGLE and IllustrisTNG simulations. Now using the CAMELS project, described in detail in \S\ref{sec:camels}, we can for the first time perform this analysis on simulations in which individual feedback parameters are varied, to study the effect on the feedback energy, the nonlinear coupling of different feedback mechanisms, and the corresponding correlation with CGM features. Crucially, the CAMELS boxes are the same size, same resolution, cosmological parameters, and initial seed, allowing for a direct comparison between subgrid models and parameter variations.

Feedback is calculated cumulatively, meaning that we add the contribution of each mode over time. In general, there are two sources of feedback in simulation subgrid models: SNe and AGN. AGN feedback is divided into several modes depending on the specifics of its implementation. The total cumulative feedback energy is then given by $E_{\rm FB,total} = E_{\rm FB, SNe} + \sum^{n}_{i=1} E_{\rm FB, AGN_i}$. 

To contextualize these feedback energies we normalize by the halo binding energy, following \cite{davies_quenching_2020}. To estimate the binding energy of the halos in each simulation, we follow a similar method to that of \cite{wellons_exploring_2023}. The halo binding energy is given by $E_{\rm b} = f_{\rm b, cosmic}M_{\rm h} V_{\rm vir}^2$,
where $M_{\rm h}$ is the halo mass, $V_{\rm vir}$ is the halo virial velocity, and the cosmic baryon fraction $f_{\rm b, cosmic}$, the fraction of matter that is baryonic, given by $f_{\rm b, cosmic}=\Omega_b/\Omega_m$, with $\Omega_b=0.049$ for the CAMELS simulations and $\Omega_m$ is determined by the parameters of the 1P set. The virial velocity of the halo is given by
$V_{\rm vir}=\sqrt{\frac{GM_{\rm vir}}{r_{\rm vir}}}$. Thus, the binding energy of the baryons in the halo is equal to:
$E_{\rm b} = (\Omega_b/\Omega_m) G M_{\rm h}^2/r_{\rm vir}$.  We use this equation to calculate the binding energy for all halos in each simulation for both SIMBA and IllustrisTNG, using $R_{200c}$ as the virial radius and $M_{200c}$ as the halo mass. Note that, we use $f_{\rm b, cosmic}$ rather than $f_{\rm b, halo} = M_{b}/M_{h}$ to calculate binding energy as it gives us an idea baseline idea of the energy we expect ignoring feedback effects which deplete halos of baryons. 

We are primarily interested in feedback energetics and their impact on how they redistribute baryons in a halo. As in \cite{davies_quenching_2020}, we examine the CGM gas fraction $f_{\rm CGM}$ as a function of halo mass and feedback energetics. In addition, we study the closure radius, a recently introduced method of defining the extent of the CGM/halo.

$f_{\rm CGM}$ is defined as the ratio of the halo mass in the CGM to the total halo mass, 
\begin{equation}
f_{\rm CGM} = M_{\rm h, CGM}/M_{\rm h, total}.
\end{equation}
As the boundaries of the CGM are uncertain we cannot exactly quantify this number but we can provide an estimate. We estimate $f_{\rm CGM}$ via a simple procedure. We roughly define the CGM mass to be comprised of all bound gas particles to the halo that are not part of the central galaxy or any other subhalo (satellites). We calculate this mass by subtracting the gas mass in subhalos, defined as the mass within two times the stellar half-mass radius of each subhalo, from the total gas mass of the halo. We take this mass to be that of $M_{\rm h,CGM}$ and divide it by the total halo mass to get $f_{\rm CGM}$. We give values of $f_{\rm CGM}$ normalized by $f_{\rm b, cosmic}$. This calculation has been validated against other methods of calculating $f_{\rm CGM}$ and is roughly consistent. While this is still an approximation, there is no fixed and consistent way to define the extent of CGM. For the purpose of our analysis, our main concern is using a consistent approximation so as to be able to compare differences across simulations with different parameter variations and different subgrid models.

Recently, \cite{aryomlou2023} introduced the closure radius, defined as the radius enclosing a cumulative baryon fraction equal to the cosmic baryon fraction. This measurement is another way to quantify the distribution of baryons within a halo and the source of the dominant feedback mechanisms. \cite{aryomlou2023} calculated the closure radius as a function of halo mass for the original versions of IllustrisTNG, SIMBA, and EAGLE, as well as some feedback variations of IllustrisTNG (i.e., no stellar feedback and no AGN feedback). We calculate for the first time the closure radius as a function of varying SNe and AGN feedback parameters in IllustrisTNG and SIMBA. To do so, we calculate the mass of all baryons within spheres of radius $r$, $M_\text{b}=M_\text{\rm gas}+M_\text{\rm BH}+M_{*}$, where $M_\text{\rm gas}$ is the mass of gas particles, $M_\text{\rm BH}$ is the mass of black hole particles, and $M_{*}$ is the mass of all star particles. We then calculate the total mass within that sphere, $M_{\rm tot}=M_{\rm b}+M_{\rm dm}$, where $M_{\rm dm}$ is the mass of dark matter particles. We take the closure radius to be the radius at which the baryon fraction within the sphere of radius $R_c$, $f_{\rm b, halo}=M_{\rm b}/M_{\rm tot}$, is equal to $f_{\rm b, cosmic}$.     

\section{Simulations}\label{sec:data}
\subsection{The CAMELS Project}\label{sec:camels}

Cosmology and Astrophysics with MachinE Learning Simulations (CAMELS)\footnote{\href{https://www.camel-simulations.org/}{https://www.camel-simulations.org/}} is the largest collection of cosmological hydrodynamical simulations (currently consisting of 14,091 simulations, including 7,928 magnetohydrodynamic simulations and 6,163 N-body simulations) and aims to improve our understanding of baryonic feedback physics effects \citep{camels_presentation, camels_data_release}. The CAMELS project is currently comprised of ten different simulation suites using a variety of codes and subgrid models: IllustrisTNG, SIMBA, Astrid, Magneticum, SWIFT-EAGLE, Ramses, Enzo, CROCODILE, Obsidian, and N-Body. The N-body suite has a corresponding dark matter-only simulation for each CAMELS hydrodynamic simulation, with the same cosmology and random seed value, run with GADGET-3 \citep{Springel_2005b}. In addition to varying the code and subgrid model used, simulations also vary cosmology and feedback physics model parameters, ultimately resulting in thousands of simulations with distinct models of baryonic feedback physics. Each CAMELS suite is divided into multiple sets, depending on the suite these can include the Latin-Hypercube (LH) set, the 1-parameter at a time (1P) set, the Cosmic Variance (CV) set, the Extreme (EX) set, the Butterfly Effect (BF) set, and the Sobol Sequence (SB) set.

In this work, we analyze a subset of simulations from the CAMELS-SIMBA and CAMELS-IllustrisTNG suites, the original two simulations in the CAMELS project. The CAMELS-SIMBA suite is run with the GIZMO code \citep{Hopkins_2015} and the same subgrid physics as the original SIMBA simulations \citep{dave_simba_2019}. The CAMELS-IllustrisTNG suite is run with the AREPO code \citep{Springel_2010, Weinberger_2020} and the same subgrid physics as the original IllustrisTNG simulations \citep{Weinberger_2017, pillepich_illustris_sim_2018}. The details of the simulations code and subgrid models are summarized in Table \ref{tab:subgrid} and described in detail in sections \ref{sec:simba} and \ref{sec:tng} for CAMELS-SIMBA and CAMELS-IllustrisTNG respectively. It is essential to note that CAMELS-SIMBA and CAMELS-IllustrisTNG differ in their implementation of gravity and hydrodynamic solvers, radiative cooling parameterization, star formation and evolution, and feedback from galactic winds and AGN and these differences can lead to significantly different predictions.

Both of these suites are extensive and comprehensive, with the CAMELS-SIMBA suite currently containing 1,171 hydrodynamic simulations and the CAMELS-IllustrisTNG suite comprising 3,219 hydrodynamic simulations. Both the original iterations of IllustrisTNG and SIMBA as well as the CAMELS versions are widely used. Focusing on these suites allows us to place our results in the context of complementary prior studies that looked at the original versions, such as \cite{wright2024} which examines inflows and outflows in SIMBA and IlustrisTNG, and studies using the CAMELS versions with parameter variations, such as \cite{Medlock_2024} which explores the potential of FRBs to constrain baryonic feedback models. A similar analysis on the since added 7 additional subgrid models of the CAMELS project is worthwhile but beyond the scope of this paper.

To analyze the impact of each parameter individually on overall feedback energetics, we use the 1P set, which in its original form consists of 61 simulations for each suite, varying six parameters. In this set, only one cosmological or astrophysical parameter is varied at a time (hence the name one parameter at a time), while the random seed values (initial conditions) remain constant. The original 1P set has two cosmological parameters ($\Omega_{\rm m}$ and $\sigma_8$) and four astrophysical parameters, which modulate stellar feedback ($A_{\rm SN1}$ and $A_{\rm SN2}$) and AGN feedback ($A_{\rm AGN1}$ and $A_{\rm AGN2}$). $\Omega_{\rm m}$ is the fraction of energy density in matter (baryonic and dark matter combined) and is linearly sampled from 0.1 to 0.5, with the fiducial value set at 0.30. $\sigma_8$ is the variance of the linear field on a scale of $8$ $h^{-1}$Mpc at $z = 0 $ and is linearly sampled from 0.6 to 1.0, with the fiducial value set at 0.80. Table \ref{tab:subgrid} summarizes key information on the four astrophysical parameters used in this study for the IllustrisTNG and SIMBA models. The parameters are implemented as prefactors to the fiducial values in the original simulations, so a value of 1.0 corresponds to the fiducial value. The physical meanings of these parameters vary depending on the suite/subgrid model. More recently, the 1P set has been extended to 28 parameters (the 1P-28 set) for CAMELS-IllustrisTNG and CAMELS-SIMBA \citep{camels_data_release2}. These parameters include the six from the original set as well as 22 new parameters, including three cosmological parameters that previously were kept constant: $\rm \Omega_{b}$, $h$, and $\rm n_s$. These parameters include 5 cosmological parameters, 14 parameters related to stellar feedback, and 9 related to AGN feedback.

In our primary analysis we focus on the four astrophysical feedback parameters included in the original 1P set. Although these four parameters do not capture all the key baryonic processes, they allow us to narrow our study to two of the most important yet not well-understood processes: galactic winds driven by supernovae (SNe) and kinetic feedback from massive black holes. In combination with previous work exploring the effect on galaxy and halo properties of varying these four parameters in CAMELS, our analysis allows us to make a connection and interpret the details of the subgrid implementation directly to feedback strength and impact. It is important to note again that CAMELS-SIMBA and CAMELS-IllustrisTNG differ significantly in subgrid implementation, so parameter comparisons are not one-to-one. However, the parameters capture the same physical processes (e.g. the velocity of galactic winds) and thus the differences that may arise between the two allow us to understand how different implemntations lead to different results. In addition, we analyze the CAMELS-IllustrisTNG 1P-28 set, whose 28 parameters are described in Table \ref{tab:1P28_desc} in the Appendix. This analysis allows us to take the next step of identifying which parameters have the most significant impact and serve as a reference for future work.

\LTcapwidth=\textwidth
\renewcommand{\arraystretch}{1.5} 
\begin{table*}
\begin{longtable*}{|>{\centering\arraybackslash}p{2.5cm}|>{\raggedright\arraybackslash}p{7.2cm}|>{\raggedright\arraybackslash}p{7.2cm}|}
\hline
 & \normalsize \textbf{CAMELS-SIMBA} & \normalsize \textbf{CAMELS-IllustrisTNG}  \\
 & \cite{dave_simba_2019, camels_presentation} & \cite{pillepich_illustris_sim_2018, camels_presentation} \\
 
\hline \hline
Hydrodynamics & GIZMO, Meshless Finite Mass \citep{Hopkins_2015} & AREPO, Moving Voronoi Mesh \citep{Springel_2010}\\
\hline
Star Formation & $H_2$ based \citep{kennicutt_sfr_1998, Krumholz_2011, krumholz_sf_2015} & $\rho$ threshold \citep{kennicutt_sfr_1998, Springel_Hernquist_2003} \\
\hline
Stellar Feedback & Thermal \& Kinetic (30\% injected hot) \newline Wind mass loading/velocity from FIRE \newline \citep{muratov_2015, daa_2017b} & Thermal \& Kinetic (90\% Momentum, 10\% Thermal) \newline Wind elements decoupled \newline \citep{pillepich_illustris_sim_2018} \\
\hline
$A_{\rm SN1}$ & Galactic winds: mass loading (wind mass outflow rate per unit SFR), prefactor of $\eta(M_*)$ in Eq.~1 of \cite{dave_simba_2019} \newline Fiducial Value: $A_{\rm SN1} = 1.0$, varied logarithmically from [0.25-4.0]  &  Galactic winds: energy per unit SFR, prefactor of $\Bar{e}_w$ in Eq.~3 of \cite{pillepich_illustris_sim_2018} \newline Fiducial value: $A_{\rm SN1} = 1.0$,  varied logarithmically from [0.25-4.0]  \\
\hline
$A_{\rm SN2}$ & Galactic winds: wind speed, prefactor of $v_w$ in Eq.~3 of \cite{dave_simba_2019} \newline Fiducial Value: $A_{\rm SN2} = 1.0$, varied logarithmically from [0.5-2.0] & Galactic winds: wind speed, prefactor of $\kappa_w$ in Eq.~1 of \cite{pillepich_illustris_sim_2018} \newline Fiducial Value: $A_{\rm SN2} = 1.0$, varied logarithmically from [0.5-2.0] \\
\hline
AGN Feedback & Accretion: Bondi-Hoyle (Hot), torque-limited (cold) \newline Low Accretion (jet mode): Kinetic \& thermal \newline High Accretion (radiative mode): Kinetic feedback \newline $M_{\rm BH}$ Threshold: $10^{7.5} M_{\odot}$  \newline \citep{angles-alcazar_gravitational_2017} &  Accretion: Modified Bondi \newline Low Accretion: Kinetic Mode \newline High Accretion: Thermal Mode \newline $M_{\rm BH}$ Threshold: $10^{8} M_{\odot}$  \newline \citep{Weinberger_2017}  \\
\hline
$A_{\rm AGN1}$ & QSO and jet-mode BH-Feedback: momentum flux, prefactor of $\dot{P}_{\rm out}$ in Eq.~10 of \cite{dave_simba_2019} \newline Fiducial Value: $A_{\rm AGN1} = 1.0$, varied logarithmically from [0.25-4.0]  & Kinetic-mode BH-Feedback: energy/unit BH accretion rate, prefactor in front of RHS in Eq.~8 of \cite{Weinberger_2017} \newline Fiducial Value: $A_{\rm AGN1} = 1.0$, varied logarithmically from [0.25-4.0]\\
\hline
$A_{\rm AGN2}$ & Jet-mode BH-Feedback: jet speed, prefactor of $v_{\rm jet}$ in Eq.~8 of \cite{dave_simba_2019} \newline Fiducial Value: $A_{\rm AGN2} = 1.0$, varied logarithmically from [0.5-2.0]  & Kinetic-mode BH-feedback: ejection speed/burstiness, prefactor of $f_{\rm re}$ in Eq.~13 of \cite{Weinberger_2017}\newline Fiducial Value: $A_{\rm AGN2} = 1.0$, varied logarithmically from [0.5-2.0] \\
\hline
\caption{Overview of the numerical methods and astrophysical subgrid models implemented in the two simulations, SIMBA and IllustrisTNG, used in the paper. We include a description of the parameters that are varied in the CAMELS project and their physical meaning for each model.}
\label{tab:subgrid}\\
\end{longtable*}
\end{table*}

All simulations follow the evolution of $\rm 256^3$ dark matter particles and fluid elements (for hydrodynamical simulations) from $\rm z = 127$ to $\rm z = 0$ in a periodic box with sides of comoving length $\rm L = 25$ $h^{-1}$ Mpc. The dark matter resolution elements are of $M_{\rm DM} = 6.49 \times 10^{7} (\Omega_{\mathrm{m}} - \Omega_{\mathrm{b}})/0.251 h^{-1} M_\odot$ and gas mass resolution is initially set to $M_{\rm b} = 1.27 \times 10^{7} h^{-1} M_\odot$. CAMELS-IllustrisTNG has mass and spatial resolution comparable to IllustrisTNG300-1, which has $\epsilon_{gas, min} = 0.25$ ckpc/h, $r_{cell, min} = 47$ pc, and $\Bar{r}_{cell} = 31.2$ kpc \citep{nelson2019illustristng}. CAMELS-IllustrisTNG uses a gravitational softening length for dark matter and stellar particles of $\epsilon_{DM,*} \sim 2$ ckpc. CAMELS-SIMBA has the same resolution as the original SIMBA 100 Mpc simulation and includes adaptive gravitational softenings for gas, stellar, and dark matter components, with the minimum gravitational softening set to 0.5 kpc/h. A majority of the simulations are performed with the following cosmological parameters: $\rm \Omega_{b} = 0.049$, $h = 0.6711$, $\rm n_s = 0.9624$, $\Sigma m_s = $0.0~eV, and $w = -1$, with the exception of a subset in the 1P-28 set for which $\rm \Omega_{b}$, $h$, and $\rm n_s$ are varied, described in the following sections and the Appendix. For more information the CAMELS project, we refer the reader to \citet{camels_data_release, camels_presentation, camels_data_release2}. 

\subsection{Halo Selection}

Halos in each simulation set are identified via a Friends-Of-Friends (FOF) algorithm and subhalos with a SUBFIND algorithm. For each simulation, we impose a minimum halo mass of greater than $\rm 10^{11} M_{\odot}$. We match black holes to subhalos by computing the Euclidean distance from each black hole to all eligible subhalos and choosing the subhalo with the minimum distance. Each halo's total cumulative feedback energy is obtained by combining stellar and AGN feedback energy calculations, as described for SIMBA in \S\ref{sec:simba} and for Illustris-TNG in \S\ref{sec:tng}.

\subsection{Feedback Implementation in SIMBA} \label{sec:simba}
SIMBA is built on the MUFASA cosmological galaxy formation simulations and GIZMO meshless finite-mass hydrodynamics \citep{dave_simba_2019}.

\subsubsection{Stellar Feedback in SIMBA}
SIMBA uses FIRE simulations \citep{hopkins_fire_2014} to quantify mass outflow rates for individual particles in a star-forming region \citep{daa_2017b}. Stellar winds are decoupled as two-phase winds, with 30\% of hot ejected wind particles. CAMELS also implements two additional factors: $A_\text{SN1}$ and $A_\text{SN2}$, which affect the galactic wind mass loading and the wind speed, respectively.

We first select a halo from the $z=0$ snapshot and track each primary subhalo through previous snapshots with the SubLink catalog. At each snapshot, we record the group halo properties: total mass ($M_{200c}$), radius ($R_{200c}$), stellar mass ($M^*$), and star formation rate (SFR), obtained through calculations of the total mass of the group enclosed in a sphere with mean density 200 times the mean density of the Universe and the corresponding radius of the sphere, the sum of the individual masses of all the star particles in the group and the sum of individual SFRs of the gas particles in the group, respectively. For each snapshot, we then calculate the rate of stellar feedback energy injection (in the form of kinetic energy) and obtain the total kinetic feedback energy by integrating across all of the snapshots. 

The wind velocities are given by
\begin{equation}
v_w=A_\text{SN2}\times1.6\left(\frac{v_\text{circ}}{200\text{km s}^{-1}}\right)^{0.12}v_\text{circ},
\end{equation}
where $v_\text{circ}=GM_{200}/R_{200}$ is the circular velocity of the halo. SIMBA simulations also include a term $\Delta v(0.25 R_\text{vir})$ that acts as a boost factor corresponding to the velocity associated with the potential difference between the launch position and the fractional virial radius \citep{muratov_2015}. \cite{dave_mufasa_2016} find that this term can increase wind speeds by significant amounts in low-mass galaxies and much less in more massive galaxies. We find that the boost factor does not add a significant difference to the overall feedback energy calculation, resulting in small additions relative to the total wind speeds, which are on the order of hundreds to thousands of km/s at the halo mass range that we use in this work. As we do not know the launch positions of each individual supernova, we thus omit the boost factor from our calculations. 

The mass loading is obtained through 
\begin{equation}
    \eta(M^*) = A_\text{SN1}\times 9\left(\frac{M_*}{M_0}\right)^{-\alpha}
    \begin{cases}
        \alpha=0.317,\hspace{0.1cm} M*<M_0 \\
        \alpha=0.761,\hspace{0.1cm} M*>M_0,
    \end{cases}
\end{equation}
where $M*$ is the stellar mass of the halo and $M_0=5.2\times10^9M_\odot$. In CAMELS, the prefactor A$_\text{SN1}$ normalizes the mass loading factor in these equations. The mass of the outflow in the winds is then given by $\dot{M}_{\rm wind}=\eta \times SFR$. The kinetic feedback energy used to generate the winds is given by $E_{\rm FB}=\frac{1}{2}\dot{M}_{\rm wind}v_w^2$, which we then integrate across the redshifts of each snapshot, ranging from $z=6$ to $z=0$. We note that this method of calculating stellar feedback energy is approximate and represents an estimate that is used quantitatively in our analysis.

\subsubsection{AGN feedback in SIMBA}
Black holes in SIMBA are seeded in galaxies with $M^* \gtrsim 10^{9.5}M_\odot$, motivated by the findings from FIRE simulations that stellar feedback suppresses black hole growth below this stellar mass threshold \citep{daa_2017c}. The black hole feedback is implemented via a kinetic subgrid model, motivated by the bipolar feedback observed in AGN outflows \citep{heckman_2014}. In addition, SIMBA has a two-mode accretion model: torque-limited accretion and Bondi accretion. \cite{hopkins_analytic_2011} find that using the torque-limited accretion model in addition to Bondi accretion results in more consistent models of black hole growth. 

The numerical implementation is fully defined in \cite{dave_simba_2019} and is summarized here. The total mass accretion rate onto a black hole is given as
\begin{equation}
    \dot{M}_{\rm BH} = (1-\eta)(\dot{M}_{\rm Bondi}+\dot{M}_{\rm Torque}),
\end{equation}
where $\eta=0.1$ is the radiative efficiency \citep{marconi_local_2004}. Accretion is limited based on the Eddington accretion rate, with three times the Eddington limit for torque-limited accretion, because it is nonspherical and accretion can exceed the Eddington limit as is observed at higher redshifts. Accretion proceeds based on the models of \cite{springel_modelling_2005}: a fraction of the mass of gas particles within a radius $R_0$ are subtracted and added to the black hole, and if a particle is sufficiently small compared to its original mass, all of it is added to the black hole \citep{thomas_black_2019}. 

The gravitational torque-limited accretion model is related to the inflow of cold gas driven by disc gravitational instabilities, and it is estimated from the following equation from analytic calculations of angular momentum transport and gas inflow in galaxies \citep{hopkins_analytic_2011,angles-alcazar_gravitational_2017}:
\begin{multline}
    \dot{M}_{\rm Torque}=\epsilon_Tf_d^{5/2}\times \left(\frac{M_{\rm BH}}{10^8M_\odot}\right)^{1/6}\left(\frac{M_{\rm enc}(R_0)}{10^9M_\odot}\right) \\
    \times \left(\frac{R_0}{100~{\rm pc}}\right)^{-3/2}\left(1+\frac{f_0}{f_{\rm gas}}\right)^{-1}M_\odot \; yr^{-1},
\end{multline}
where $f_{\rm d} = M_d(R_0)/[M_{\rm gas}(R_0)+M_*(R_0)]$ is the disc mass fraction including stars and gas, $M_{\rm enc}$ is the total gas and stellar mass, $f_{\rm gas}=M_{\rm gas}(R_0)/M_d(R_0)$ is the gas mass fraction in the disc component, and $f_0 \approx 0.31 f_{\rm d}^2(M_d(R_0)/10^9M_\odot)^{-1/3}$.
Quantities are evaluated within a distance of $R_0$ of each black hole enclosing the nearest 256 gas elements \citep{dave_simba_2019}. $\epsilon_T = \epsilon_m\times \alpha_T$ is the normalization factor that encapsulates processes affecting the radial transport of gas on unresolved scales (e.g., mass loss in the accretion disk winds and stellar feedback) where $\alpha_T=5$ is the normalization factor corresponding to the efficiency of transport of material from the inner galactic disc onto the black hole accretion disk and $\epsilon_m=0.1$ is the additional normalization that accounts for unresolved scales and corresponds to the efficiency of transport from the accretion disc onto the black hole and is set to 10\% as the canonical value set to match the local $M_{\rm BH}-M^*$ relation. The value of the total normalization factor is chosen to match the observed $z=0$ scaling relations \citep{angles-alcazar_black_2013, angles-alcazar_torque-limited_2015}.

For hot gas ($T>10^5$K), the Bondi parameterization is used, as hot gas is more spherically distributed, and this accretion mode is subdominant for all but the highest mass black holes. It is derived analytically in \cite{bondi_spherically_1952} as

\begin{equation}    \dot{M}_{\rm Bondi}=\epsilon_m\frac{4\pi G^2M^2_{\rm BH}\rho}{(v^2+c_s^2)^{3/2}},
\end{equation}
where $\rho$ is the mean density of hot gas within the black hole accretion kernel, $c_s$ is the kernel-averaged sound speed of the gas, $v$ is the kernel-averaged velocity of the gas relative to the black hole, and $\epsilon_m=0.1$ is the suppression factor that compensates for high gas temperatures due to the multiphase subgrid model of star formation \citep{thomas_black_2019}.

SIMBA uses a kinetic subgrid model for black hole feedback that mimics energy injection into the surrounding gas using bipolar feedback with velocities and temperatures determined from observations of AGN outflows. There are three modes of black hole feedback: radiative, jet, and X-ray heating. In this work, we focus on the former two modes of kinetic feedback.

The radiative mode is present at high Eddington $f_{\rm Edd}=\dot{M}_{\rm BH}/\dot{M}_{\rm Edd}>$ a few percent. During this phase, AGNs drive multiphase winds of warm ionized gas at $v\sim1000$ km/s with the outflow velocity of 
\begin{equation}
    v_{\rm w,EL} = 500 + 500(\log M_{\rm BH} - 6)/3 \text{ km s}^{-1},
\end{equation}
based on ionized gas linewidth observations of X-ray detected AGN from SDSS \citep{perna_x-raysdss_2017}.

The jet mode is used at low Eddington ratios ($f_{\rm Edd}<0.2$), during which AGNs drive hot gas in collimated jets at $v\sim10^4$ km/s forming hot bubbles in the surrounding gas. The jet outflow velocity is given as
\begin{equation}
    v_{\rm w,jet}=v_{\rm w,EL}+7000\log(0.2/f_{\rm Edd})\text{ km s}^{-1},
\end{equation}
motivated by observations of hot gas ejected at high velocities, on the order of $10^4$ km/s \citep{fabian_2012}. The jet mode is triggered as $M_{\rm BH}>M_{\rm BH,lim}=10^{7.5}M_\odot$, motivated by observations from \cite{barisic_stellar_2017}, who find that radio jets occur only in systems with high galaxy velocity dispersions corresponding to $M_{\rm BH}>10^8M_\odot$.

The CAMELS simulations implement two additional parameters, AGN1, which vary the total momentum flux, and AGN2, which varies the maximum jet velocity, with the outflow velocity given by

\begin{equation}
    v_\text{out} = 
    \begin{cases}
        v_\text{rad}+A_\text{AGN2}\times v_\text{jet} \text{\hspace{0.43cm} if} \begin{matrix}
            & \lambda_\text{\rm Edd}<0.2 \text{ \&}\\ 
             & M_\text{BH}>10^{7.5}M_\odot
        \end{matrix}\\
            
        v_\text{rad} \hspace{3cm} \text{otherwise}
    \end{cases}
\end{equation}
The amount of material ejected in the outflow winds due to AGN feedback is based on observations of AGN outflows energetics, such as momentum and energy ratios \citep{fiore2017,ishibashi2018}. The mass loading factor in AGN winds scales inversely with the outflow velocity, and the outflows are ejected in bipolar directions, parallel or antiparallel to the angular momentum vector of the gas within $\rm{R}_0$, the distance enclosing the nearest 256 gas particles to the black hole, as described in \cite{dave_simba_2019}. The mass loading factor is determined by the momentum input $\dot{P}_{\rm out}=\eta\dot{M}_{\rm BH}c^2$ and the bolometric luminosity of the AGN ($L_{\rm bol}$):

\begin{equation}
\dot{P}_{\rm out}=\dot{M}_{\rm out}v_{\rm out}=A_{\rm AGN1}\times \frac{20L_{\rm bol}}{c}.
\end{equation}

\noindent The cumulative energy injected into the surrounding medium by black hole feedback is calculated by tracking each black hole over its lifetime and calculating the energy from radiative and jet phase outflows at each time step.

\subsection{Feedback Implementation in IllustrisTNG} \label{sec:tng}

The CAMELS-IllustrisTNG suite employs the galaxy formation model of the IllustrisTNG simulations, described in \citet{Weinberger_2017, pillepich_illustris_sim_2018} which builds upon its predecessor Illustris (described in \citet{Vogelsberger_2013} and \citet{Torrey_2014}). In this section, we will give a brief overview of stellar and AGN feedback implementations in IllustrisTNG to provide context and a physical understanding of our results.

\subsubsection{Stellar Feedback in IllustrisTNG}
The galactic wind scheme is based on \citet{Springel_2003}, in which the winds driven by stellar feedback are implemented kinetically with temporarily hydrodynamically decoupled particles that are then stochastically and isotropically (initial velocities in random directions) ejected from the star-forming gas particles. In addition, there is a subdominant thermal energy component ($10\%$ in the fiducial model). Two key quantities, the wind speed $v_w$ and the total energy injection rate per unit star formation $e_w$ depend on physical properties of the local medium (including metallicity and dark matter velocity dispersion), redshift, and two global normalization parameters. The CAMELS-IllustrisTNG suite introduces two additional parameters to modulate these two quantities. Wind speed is given by,
\begin{equation}
    v_w = A_{\rm SN2} \times \text{max}\left[\kappa_w \sigma_{DM} \left(\frac{H_0}{H(z)}\right)^{1/3}, v_{w,\rm min}\right],
\end{equation}
\noindent where $\kappa_w$ is the wind velocity factor, a multiplicative dimensionless parameter (set to 7.4 for the fiducial model), $\sigma_{DM}$ is the 1D dark matter velocity dispersion, measured with a weighted kernel over the $N=64$ nearest particles, $H_0$ is the Hubble constant, and $H(z)$ is the Hubble parameter. A minimum possible $v_w$ is imposed as $v_{\rm w, min}$ (set to 350 km/s for the fiducial model). This minimum is implemented to ensure that the mass loading factors in low-mass halos do not become unphysically high. CAMELS introduces the additional multiplicative parameter $A_{\rm SN2}$, which is set to 1 for the fiducial model, and varied logarithmically from 0.5 to 2. The 1P-28 set additionally introduces variations in $v_{w,\rm min}$.

The mass loading factor $\eta_w$ is calculated using the wind velocity $v_w$ and the total energy injection rate per unit star formation $e_w$ with the equation,
\begin{equation}
    \eta_w = \frac{\Dot{M}_w}{\Dot{M}_{\rm SFR}} = \frac{2}{v_w^2} e_w (1 - \tau_w),
\end{equation}
where $\Dot{M}_w$ is the rate of gas mass being converted to wind particles, $\Dot{M}_{\rm SFR}$ is the instantaneous local star formation rate, and $\tau_w$ is the fraction of the available wind energy that is thermal (varied in 1P-28). 

The energy available in a star-forming gas cell with metallicity $Z$ is parameterized similarly to that of EAGLE \citep{Schaye_2015} and is described by,
\begin{equation} \label{eq:pil_3}
    e_w = A_{\rm SN1} \times \omega \times N_{\rm SNII} \mathbf{E_{\rm SNII,51}} \hspace{0.1cm} M_{\odot}^{-1},
\end{equation}
with $\omega = \Bar{e}_w \left[f_{\rm w,Z} + \frac{1 - f_{\rm w,Z}}{1 + (Z/Z_{\rm w,ref})^{\gamma_{\rm w,Z}}}\right]$, where $\Bar{e}_w$ is the wind energy factor, a dimensionless free parameter of the model, $f_{\rm w,Z}$ is the metallicity dependence reduction factor, $Z_{\rm w,ref}$ is the metallicity dependence reference metallicity, $\gamma_{\rm w,Z}$ is the metallicity dependence reduction power, $N_{\rm SNII}$ is the number of SNII per formed stellar mass in solar mass units, and $E_{\rm SNII,51}$ is the available energy per core-collapse supernova in units of $10^{51}$ erg. CAMELS introduces the extra multiplicative parameter $A_{\rm SN1}$, set to 1 to match the fiducial model, and it varied logarithmically from 0.25 to 4. The 1P-28 set also varies the parameters: $f_{\rm w,Z}$, $Z_{\rm w,ref}$, and $\gamma_{\rm w,Z}$. 

The metallicity dependence implies,
\begin{equation*}
    \eta_w \propto f_{\rm w,Z} \times \frac{2}{v^2_w} \Bar{e}_w (1 - \tau_w) 
\end{equation*}
where $f_{w,Z} = 1$ when $z \ll Z_{\rm w,ref}$.
In essence, this means that the injection energy for gas cells with metallicities that are much larger than the set threshold of $Z_{\rm w,ref}$ is reduced by the factor $f_{\rm w,Z}$. Thus, galactic winds are weaker in higher-metallicity environments. This is motivated by the idea that higher metallicity galaxies may have larger radiative cooling losses of the SN energy \citep{martizzi_2015, Schaye_2015}. In addition, the metallicity dependence of the evolution, mass loss, and end state of SNII progenitors \citep{smartt_2009} affects the amount of energy released into the ISM per core collapse. 

In the fiducial model, the parameters are set as follows: $\Bar{e}_w = 3.6$, $f_{\rm w,Z}= 0.25$, $Z_{\rm w,ref} = 0.002$, $\gamma_{\rm w,Z} = 2$, and $E_{\rm SNII,51} = 1$. The wind energy is, in practice, then given by Eq.~\ref{eq:pil_3} with $\omega$ varying from $[3.6, 0.9]$, according to increasing gas metallicity. For typical MW metalicities of about $Z = 0.015 - 0.02$, this results in galactic winds that are launched with $e_w \sim N_{\rm SNII} \times 10^{51}  \hspace{0.1 cm} \text{erg}  \hspace{0.1cm} M_{\odot}^{-1}$.

The thermal energy component prevents spurious star formation and other artifacts of hydrodynamically recoupling of the wind particles. Wind particles are allowed to radiatively cool as they travel, adopting the effective density of the nearest gas cell. Winds recouple with the gas cell they are currently in either when it falls below a certain density (0.05 times the density threshold for star formation) or when the maximum travel time has elapsed (0.025 x the current Hubble time).

Overall, the total energy release rate available to drive galactic winds is given by,
\begin{equation}
    \label{eq:TNG_SN}
    \Dot{E}_w = \frac{\Dot{M}_w v_w^2}{2 (1 - \tau_w)} = e_w \Dot{M}_{\rm SFR},
\end{equation}
where all terms are spatially and time-dependent. We compute the cumulative stellar feedback using Eq.~\ref{eq:TNG_SN}. First, we use Eq.~\ref{eq:pil_3} to calculate $e_w$ with the modulated parameter values for each simulation. We then multiply $e_w$ by the initial mass of all stellar particles in a given halo as a proxy for the instantaneous SFR. This provides us with an estimate of the total cumulative energy available to drive the galactic winds.

\subsubsection{Black Holes and AGN feedback in IllustrisTNG}

Black holes are seeded with seed mass $M_{\rm seed}$ when the FOF halo finder identifies a halo more massive than the threshold mass, $M_{\rm FOF} = 5 \times 10^{10} h^{-1} M_{\odot}$, that does not already have a black hole. In the fiducial IllustrisTNG model this seed mass is $M_{\rm seed} = 8 \times 10^5 h^{-1} M_{\odot}$. In CAMELS-TNG 1P28 set this is another varied parameter.

Black holes accrete at the Bondi rate, with an upper limit on accretion set by the Eddington rate:
\begin{equation}
	\Dot{M}_{\rm BH} = \min(\Dot{M}_{\rm Bondi},\Dot{M}_{\rm Edd}).
\end{equation}
The Bondi accretion rate is given by,
\begin{equation}
	\Dot{M}_{\rm Bondi} = \frac{4 \pi G^2 M_{\rm BH}^2 \rho}{c_s^3},
\end{equation}
and the Eddington accretion rate by,
\begin{equation}
	\Dot{M}_{\rm Edd} = \frac{4 \pi G M_{\rm BH} m_p}{\epsilon_r \sigma_T c},
\end{equation}
where $G$ is the gravitational constant, $c$ is the vacuum speed of light, $m_p$ is the proton mass, $\sigma_T$ is the Thompson cross-section, $M_{\rm BH}$ is the black hole mass, and $\rho$ and $c_s$ are the density and sound speed of the gas in the vicinity of the black hole. In the Eddington accretion rate, $\epsilon_r$ is a radiative accretion efficiency, set to $0.2$ in the fiducial model.

Black holes have two possible accretion states in IllustrisTNG: a high-accretion-rate mode with classic disc accretion, and a low-accretion-rate mode with a spherical, hotter accretion flow, referred to as the 'quasar' and 'radio' modes, respectively. The Eddington ratio determines which accretion mode the black hole is operating with. Specifically, SMBHs are in the high-accretion state if their Bondi-Hoyle-Lyttleton accretion rate $\Dot{M}_{\rm Bondi}$ is greater than some fraction $\chi$ of the Eddington accretion rate $\Dot{M}_{\rm Edd}$,
\begin{equation}
	\frac{\Dot{M}_{\rm Bondi}}{\Dot{M}_{\rm Edd}} \geq \chi,
\end{equation}
where the fraction $\chi$ is expected to plausibly fall between values of $\sim 0.001 -0.1$ \citep{Dunn_2010} and has previously been implemented as a constant in simulations (ex. $\chi = 0.05$ in Illustris). In IllustrisTNG, the fraction $\chi$ is given by the black hole mass dependent scaling,
\begin{equation}
	\chi = \min\left[\chi_0 \left(\frac{M_{\rm BH}}{10^8 M_{\rm \odot}}\right)^{\beta}, 0.1\right],
\end{equation}
where the parameters $\chi_0 = 0.002$ and $\beta = 2.0$ in the fiducial model and are modulated in the CAMELS-TNG 1P28 set. This mass scaling is motivated by the fact that low-z black holes in massive systems show evidence of having transitioned to the low-accretion state. The mass dependent scaling favors this transition the low-accretion state for the most massive black holes at late times.

IllustrisTNG has three modes of AGN feedback: thermal, kinetic, and radiative. Two of these modes follow the modes of high-accretion and low-accretion rate. The thermal mode corresponds to the high accretion mode, applying to black holes with a mass less than $10^8 M_{\odot}$ and the kinetic mode to the low accretion mode, for black holes with a mass greater than $10^8 M_{\odot}$. The third mode, the radiative feedback mode, is always turned on and contributes as additional radiation flux to the cosmic ionizing background.

For the high-accretion state, the thermal feedback energy is calculated as
\begin{equation}
	\Delta \Dot{E}_{\rm high} = \epsilon_{\rm f,high} \epsilon_{r} \Dot{M}_{\rm BH} c^2,
\end{equation}
where $\Dot{M}_{\rm BH}$ is the estimated black hole mass accretion rate of the black hole with mass $M_{\rm BH}$, $\epsilon_r$ is the radiative efficiency (the fraction of the accreted rest mass energy that is released in the accretion process), and $\epsilon_{\rm f,high}$ is the fraction of this energy that couples to the surrounding gas. The thermal mode is prescribed as the injection of thermal energy spherically in a small sphere defined to hold a fixed amount of mass around the SMBH. The injection rate is equal to 0.02 times the black hole mass accretion rate. 

In CAMELS-IllustrisTNG, it is the kinetic feedback mode, corresponding with the low-accretion state, that is modulated by the two AGN feedback parameters. The kinetic feedback energy is given by
\begin{equation}
    \Dot{E}_{\rm low}= A_{\rm AGN1} \epsilon_{\rm f,kin} \Dot{M}_{\rm BH}c^2,
\end{equation}
where $\epsilon_{\rm f,kin} = \min\left(\frac{\rho}{f_{\rm th} \rho_{\rm SF,th}}, 0.2\right)$. We adopt the fiducial value of $f_{\rm th} = 0.05$ The kinetic feedback mode is implemented purely kinetically, injecting only momentum, without immediate thermal energy. The energy is injected into a sphere defined to hold a fixed amount of mass around the SMBH but not necessarily spherically, but in a random direction. 

For an available kinetic feedback energy of $\Delta E$ each gas cell j in the feedback region is kicked by:
\begin{equation}
	\Delta p_j = m_j \sqrt{\frac{2 \Delta E w(r_j)}{\rho}} n.
\end{equation}
Thus, the total momentum injection per feedback event, summing over each gas cell j in the feedback region, is given by:
\begin{equation}
	p_{\rm inj} = \sum_j m_j  \sqrt{\frac{2 \delta E w(r_j)}{\rho}} n.
\end{equation}
The corresponding change in the total energy is then given by:
\begin{equation}
	E_{\rm inj} = \delta E + \sum (p_j \cdot n) \sqrt{\frac{2 \delta E w(r_j)}{\rho}},
\end{equation}
where $p_j$ is the momentum of cell $j$ before energy injection.

Kinetic feedback energy injection is discretized by imposing a minimum accumulated energy before the feedback is released, so that energy injection is time-step independent and each event is sufficiently powerful. The kinetic feedback threshold is parameterized by 
\begin{equation}
    E_{\rm inj,min} = A_{\rm AGN2} \times f_{\rm re}\frac{1}{2}\sigma^2_{\rm DM}m_{\rm enc}.
\end{equation}
where $\sigma_{\rm DM}$ is the 1D dark matter velocity dispersion, $m_{\rm enc}$ is the gas mass enclosed in the feedback region, and $f_{\rm re}$ is a free parameter that describes the burstiness and frequency of the kinetic feedback.

In IllustrisTNG, the cumulative injection of AGN feedback from the thermal and kinetic modes is saved with each snapshot. We take these quantities directly from the snapshot.

\section{Results} \label{sec:results}
\subsection{Feedback modes and simulation differences}

\begin{figure*}
    \centering
    \includegraphics[width=\linewidth]{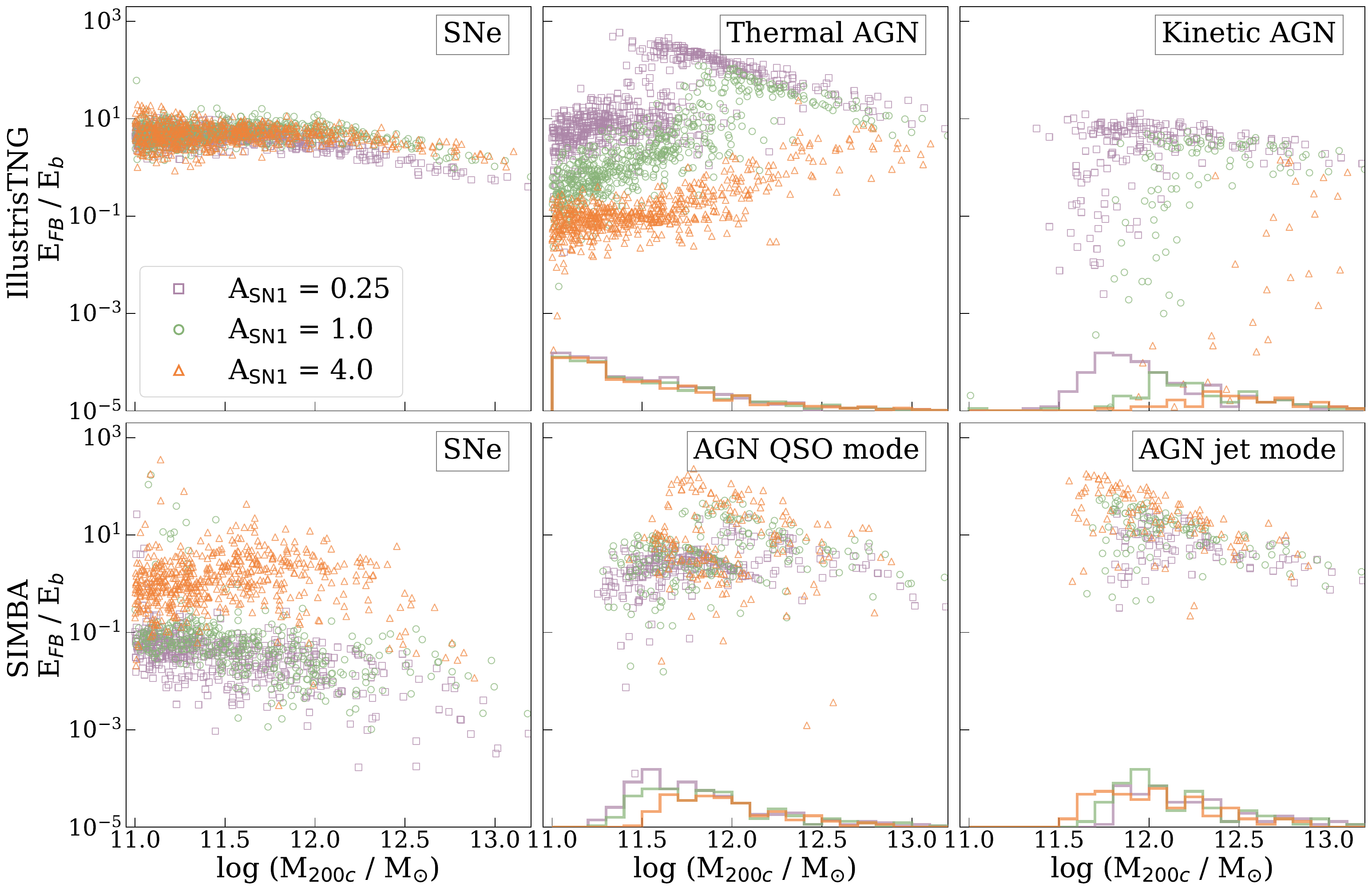}
    \caption{Cumulative feedback energy normalized by the binding energy of baryons as a function of halo mass, decomposed into the three different modes of feedback implemented in IllustrisTNG (top) and SIMBA (bottom). The first mode is feedback from supernovae; AGN feedback has two modes in IllustrisTNG (SIMBA): thermal (radiative/QSO) mode and kinetic (jet) mode. Different colored points correspond to variations in the A$_\text{SN1}$ parameter in CAMELS 1P set simulations, which controls mass loading from galactic winds in SIMBA and energy per unit SFR in IllustrisTNG. This parameter has a larger effect on the thermal AGN feedback mode in IllustrisTNG and on the SNe feedback mode in SIMBA. An increase in this parameter leads to an increase in halo mass required for AGN feedback to turn on in SIMBA and for the kinetic AGN (low accretion) mode to turn on in IllustrisTNG, as shown in the distribution of halo masses with black holes active in the corresponding feedback mode at the bottom of these panels.}
    \label{fig:asn1_decompose}
\end{figure*}

We start by examining the three feedback modes in each simulation, followed by a general comparison of SIMBA and IllustrisTNG. In IllustrisTNG, the feedback mechanisms consist of thermal and kinetic active galactic nucleus (AGN) modes, alongside a supernova (SNe) mode. Meanwhile, in SIMBA, the modes are radiative (QSO), jet AGN, and SNe. As described in \S\ref{sec:simba} and \S\ref{sec:tng}, the implementation details differ between SIMBA and IllustrisTNG for each of these modes. To analyze the separate contributions of each mode, Figure~\ref{fig:asn1_decompose} presents the feedback energy, normalized by the binding energy, as a function of the halo mass at $z=0$, separated by mode for IllustrisTNG (top row) and SIMBA (bottom row). The CAMELS project allows us to investigate, for the first time, how variations in the four feedback parameters mentioned in \S\ref{tab:subgrid} influence each individual feedback mode. In Figure~\ref{fig:asn1_decompose}, we illustrate the impact of altering one of these parameters, $A_\text{SN1}$, which affects the mass loading of galactic winds in SIMBA and the energy per unit SFR in IllustrisTNG. We show the feedback energy relative to the halo mass for low ($A_\text{SN1}$=0.25; purple), standard ($A_\text{SN1}$=1.0; green), and high ($A_\text{SN1}$=4.0; orange) values of this parameter.

First, we concentrate on the halos from the fiducial runs, plotted in green. In the cases of AGN jet and kinetic feedback modes, a distinct transition mass is observed above which these modes begin to have a significant impact. This is because the AGN feedback modes activate only at higher masses when a black hole reaches the mass threshold for the associated low-accretion-rate (high Eddington ratio) mode. In the SIMBA model, owing to the stellar mass threshold for black hole seeding and the mass threshold for activating the jet mode, the AGN feedback becomes significant only at $M_h \gtrsim 10^{11.3}M_\odot$. The AGN modes dominate the feedback budget in SIMBA with $E_{\rm FB}/E_{\rm b} \sim 1-100$ during the activation of the QSO and jet modes. SNe feedback, on the other hand, has an energy budget of $E_{\rm FB}/E_{\rm b} \sim 0.05-0.5$ and becomes a dominant feedback energy source at the low mass end ($M_h \lesssim 10^{11.5} M_{\odot}$).

In the IllustrisTNG model, the thermal AGN mode is activated for all mass scales, whereas the kinetic AGN mode turns on only when the black hole mass reaches $M_{\rm BH} \sim 10^8 \rm M_{\odot}$, which corresponds to a halo mass of at least $10^{11.3} \rm M_{\odot}$. Figure~\ref{fig:asn1_decompose} shows that at higher masses ($M_h \gtrsim 10^{11.8} M_{\odot}$), the thermal mode predominantly provides the feedback energy, with $E_{\rm FB}/E_{\rm b}$ ratios ranging from approximately 10 to 500. In contrast, both the kinetic AGN and SNe modes supply nearly equivalent levels of energy, with $E_{\rm FB}/E_{\rm b}$ ratios around 1 to 10. At lower masses ($M_h \lesssim 10^{11.8} M_{\odot}$), SNe energy predominates, with $E_{\rm FB}/E_{\rm b}$ ratios also around 1 to 10. In this lower mass range, the kinetic AGN feedback is inactive, and the thermal AGN feedback is notably weaker, with $E_{\rm FB}/E_{\rm b}$ ratios between 0.1 and 10. These findings are consistent with those reported by \cite{davies_quenching_2020} for the IllustrisTNG-100 simulation. When comparing IllustrisTNG to SIMBA, it is observed that SNe feedback in SIMBA contributes less to the overall feedback energy than in IllustrisTNG.

\begin{figure*}
    \centering
    \includegraphics[width=\linewidth]{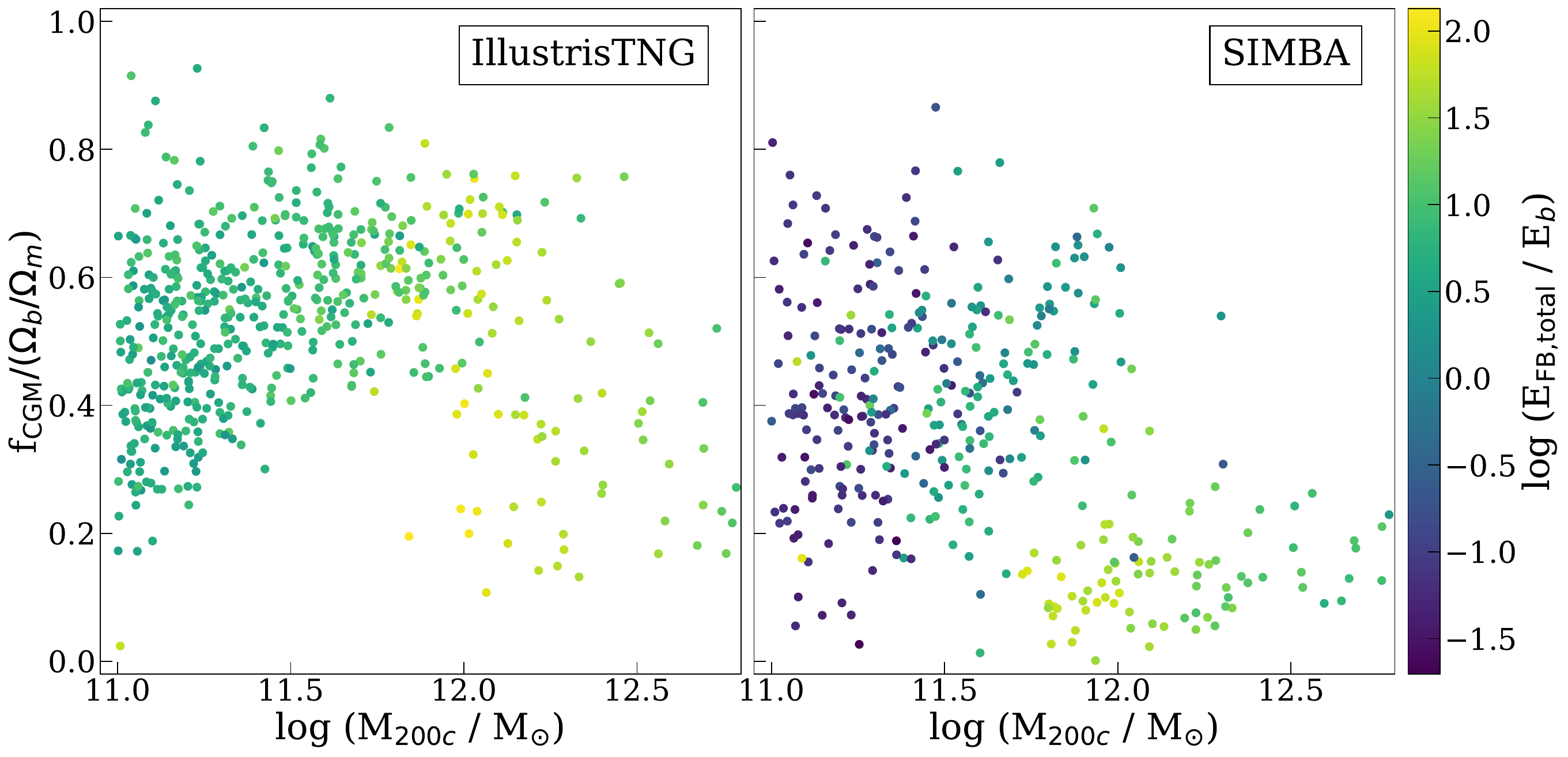}
    \caption{Normalized CGM gas fraction as a function of halo mass at $z=0$, with symbols colored by the normalized cumulative injected feedback energy over each halo's lifetime, for the fiducial runs of IllustrisTNG (left) and SIMBA (right). For both simulations, CGM gas fractions decrease as halo mass and total feedback energy relative to binding energy increases. Overall, IllustrisTNG halos have greater feedback energy than those in SIMBA with less depleted CGMs, indicating that the SIMBA feedback implementation is more efficient at moving baryons.}   \label{fig:fiducial_simba_tng}
\end{figure*}

In IllustrisTNG, increasing the energy per unit SFR ($A_{\rm SN1}$) has remarkably little effect on the SNe feedback energy alone.  The near constancy of $E_{\rm FB}$ despite a factor of $16\times$ variation of $A_\text{SN1}$ suggests that the feedback energy per unit star formation regulates the growth of the stellar component of the galaxy.  We note only a small deviation from high-mass, where a lower value of $A_\text{SN1}$ results in a slightly decreased $E_{\rm FB}$. However, the most dramatic difference appears in the energetics of AGN feedback, where an increased value of $A_\text{SN1}$ leads to a suppression of thermal and kinetic AGN feedback. On the other hand, the low value of $A_\text{SN1}$ leads to enhanced feedback energetics and more clearly highlights the threshold for which there is a jump in AGN feedback energy at $M_h \sim 3 \times 10^{11} \rm M_{\odot}$. This is because turning up $A_\text{SN1}$ suppresses the growth of black holes relative to the total halo mass, we see that this is reflected in the kinetic AGN feedback energetics. That is, as $A_\text {SN1}$ is tuned higher, the halo mass threshold for which the kinetic mode is turned on is driven higher.

In SIMBA, the most significant variation is observed in the energy output from SNe feedback. Similarly to the IllustrisTNG's kinetic mode, increasing $A_\text{SN1}$ raises the halo mass threshold for activating the AGN QSO mode by approximately 0.1~dex. Consequently, we infer that a higher mass loading factor from SNe suppresses black hole growth, delaying the activation of AGN feedback in these low-mass halos. Interestingly, the pattern evident in IllustrisTNG, where thermal AGN intensifies as $A_\text{SN1}$ decreases, is reversed in SIMBA: a higher $A_\text{SN1}$ leads to enhanced energy output in the radiative and jet AGN modes. This suggests that, in SIMBA, more efficient stellar feedback is correlated with stronger AGN feedback, whereas in IllustrisTNG, more efficient stellar feedback is correlated with weaker AGN feedback.

To gain a deeper understanding of this seemingly counterintuitive trend, we examine the black hole mass function at $z=0$ in more detail. According to \cite{camels_data_release2}, in IllustrisTNG, increasing $A_\text{SN1}$ suppresses the growth of black holes with $M_{\rm BH} \gtrsim 10^{8} M_{\odot}$, thereby delaying the onset of the kinetic AGN feedback mode. Conversely, in the SIMBA simulation, an increase in $A_\text{SN1}$ reduces the number of black holes with $M_{\rm BH} \lesssim 10^8 M_{\odot}$, while increasing those with $M_{\rm BH}\gtrsim 10^8 M_{\odot}$. In SIMBA, black holes are seeded based on a stellar mass threshold of $M_* > 10^{9.5} M_{\odot}$ \citep{dave_simba_2019, thomas_black_2019, habouzit_supermassive_2021}, which is influenced by SNe feedback, whereas in IllustrisTNG, seeding depends only on halo mass. Thus increasing $A_{\rm SN1}$ inhibits the stellar mass growth of halos in SIMBA, delaying the seeding of black holes and suppressing the number of low-mass black holes. It remains unclear why the number of high-mass black holes is enhanced. Oh et al. (in prep) find that at fixed stellar mass higher values of $A_{SN1}$ lead to higher mass black holes, hinting that perhaps increasing $A_{SN1}$ increases the growth rate of black holes. However, the physical mechanism that could be behind this remains unclear and to fully understand this trend, further analysis and study are needed.

\subsection{Feedback Energetics and $f_{\rm CGM}$}

Now, we analyze how the feedback energetics relate to the halo mass and the CGM mass fraction. Figure~\ref{fig:fiducial_simba_tng} illustrates the CGM mass fraction versus halo mass, with colors representing the normalized cumulative feedback energy injected throughout each halo's lifetime for the fiducial models at $z=0$ from the CAMELS 1P simulations. A clear demarcation between low- and high-mass halos, indicating the onset of significant AGN feedback, is evident in the plot. As the halo mass increases, the total cumulative feedback energy surpasses the binding energy, with noticeable shifts at $M_h \sim 10^{11.8} M_{\odot}$ in IllustrisTNG and $M_h \sim 10^{11.6} M_{\odot}$ in SIMBA. The CGM gas fraction diminishes in high-mass halos, correlating with an increase in cumulative feedback energy. Our results are consistent with other studies such as \cite{wright2024}, showing that CGM in SIMBA is more depleted across all mass ranges compared to IllustrisTNG, with a marked decrease in CGM gas fraction at the high-mass end due to strong AGN feedback.

Comparing SIMBA and IllustrisTNG we see that overall IllustrisTNG displays greater cumulative feedback energy (on average 0.7 dex greater) over all the halos compared to SIMBA. This is surprising as in prior work, SIMBA has been considered to have ''stronger feedback'' based on halo properties such as a depleted CGM. In this work for the first time we directly compare the feedback energy itself, which demonstrates that understanding feedback impact is not solely dependent on the total energy. Indeed, we find that SIMBA halos have a lower CGM gas fraction. This is in agreement with what other CAMELS investigations have found, that in SIMBA baryons on average spread further than in IllustrisTNG \citep{Gebhardt_2024} and are more uniformly distributed \citep{Medlock_2024}. Again, in past work this is attributed to SIMBA having ''stronger'' feedback which as Figures~\ref{fig:asn1_decompose} and \ref{fig:fiducial_simba_tng} show is clearly not the case in terms of the total cumulative feedback energy. When we do not normalize the cumulative feedback energy by the binding energies of the halos, we find the same trend. Despite this, baryons are being pushed further out from halos, indicating that even though in SIMBA feedback overall is not stronger, it more effectively couples to baryons - and is more efficient. 

To gain a clearer understanding, we examine the findings of \cite{wright2024}, who measured the inflow and outflow rates at $z = 0$ for both the original SIMBA and IllustrisTNG simulations. These results are complementary to our own in that while we focus directly on the feedback energetics, the inflow and outflow rates at the three different scales can be combined with our closure radius calculations to understand how effectively that energy couples to baryons for the fiducial iterations. \cite{wright2024} finds that at $z = 0$, SIMBA outflows can entrain CGM mass within halos where $M_{\rm h} > 10^{12} \rm M_{\odot}$. In fact, for halos with $M_{\rm h} = 10^{12.5} \rm M_{\odot}$, the rate of gas outflow on the IGM scale ($2.5 \times R_{200}$) is 5 to 10 times greater than on the ISM scale ($0.25 \times R_{200}$). This entrainment occurs because the outflows are significantly more pressurized compared to the surrounding CGM. At these halo masses, the CGM in SIMBA has low density, while the outflows themselves are thermally heated and highly pressurized by AGN feedback. Conversely, in IllustrisTNG, outflows tend to stall as they extend to greater distances, particularly in the lower mass range ($M_{\rm h} < 10^{12} \rm M_{\odot}$). In IllustrisTNG, the CGM gas densities are relatively high, and the outflowing material is underpressurized compared to the CGM and unable to entrain mass. Lastly, this can also be considered in terms of inflow rates or preventative feedback. \cite{wright2024} note that in IllustrisTNG, gas accretion rates are significantly higher than in SIMBA, up to a factor of 5 times greater for some scales.

To understand this further, it is important to contextualize it in terms of how the feedback energy is subsequently coupled to the surrounding baryons. We now turn to the details of the feedback implementation to understand why SIMBA can push baryons further than IllustrisTNG. Both SIMBA and IllustrisTNG have a kinetic AGN mode which is highly effective at moving baryons in a spatial sense. EAGLE for example has no kinetic mode, just a thermal mode, and has the lowest spatial reach compared to SIMBA and IllustrisTNG \citep{wright2024}. In general, kinetic modes are more efficient in moving baryons than thermal modes, where energy can be radiated away. The strongest feedback mode in IllustrisTNG is the thermal mode, which, as described in \S\ref{sec:tng}, deposits thermal energy spherically in a small region surrounding the SMBH. The thermal mode adds a floor to the overall $E_{\rm FB}/E_{\rm b}$ of $\sim 0.1-10$, depending on halo mass, but this energy is ineffective at removing baryons from the CGM as heated gas re-cools rapidly, as found also in \citet{davies_quenching_2020} for IllustrisTNG. The kinetic mode is subdominant and only turns on at a set black hole mass threshold, pushing and injecting energy randomly within a sphere. 

Conversely, in the SIMBA simulation, both the kinetic/jet mode and the QSO mode contribute energy through bipolar outflows. As illustrated in Figure~\ref{fig:asn1_decompose}, in SIMBA, the jet mode is slightly dominant over the QSO mode (which is also kinetic) and exhibits a slightly higher feedback energy compared to the IllustrisTNG kinetic AGN mode. The bipolar feedback implementation can more effectively propel baryons further out, since the energy is concentrated in a smaller region, making it easier to overcome the localized inflow. This contrasts with the spherical energy injection, where the outflows are fighting against the full inflow onto the halo. In essence, SIMBA employs modes that are more efficient in moving baryons in a spatial sense and exhibits greater feedback energy in those modes compared to IllustrisTNG. These features and details explain why, although the pure energetics of the IllustrisTNG feedback is greater, the SIMBA feedback is more efficient in redistributing the baryons.

\subsection{Feedback Parameter Variation Effects}

Motivated by the complex effects and interactions that we observed in Figure~\ref{fig:fiducial_simba_tng}, we now move on to an exploration of the effects of varying each of the four parameters modulating feedback energetics in the CAMELS simulations. In this section, we examine the effect on the total feedback energy, rather than individual feedback modes. We build on this analysis by studying the subsequent effect on the key properties describing the distribution of baryons in the halos: the CGM gas fraction and the closure radius, all functions of the halo mass.

\begin{figure*}
    \centering
    \includegraphics[width=\linewidth]{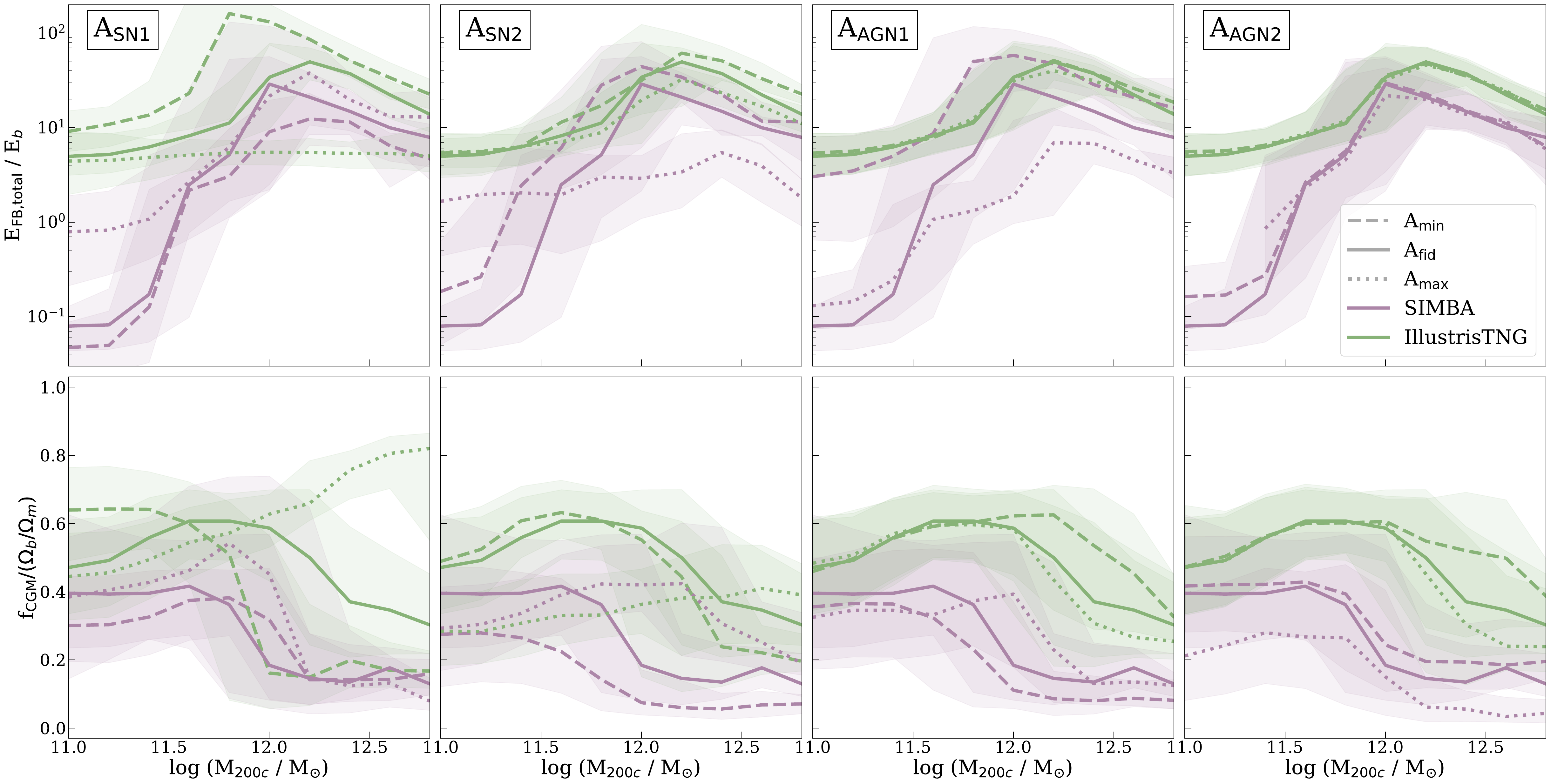}
    \caption{Top row: Total cumulative feedback energy (normalized by halo binding energy) as a function of halo mass at $z = 0$, shown as running medians. Bottom row: CGM gas fraction $f_{\rm CGM}$ (normalized by $f_{\rm b, cosmic} = \Omega_b/\Omega_m$) as a function of halo mass at $z = 0$, shown as running medians. IllustrisTNG simulations are in green and SIMBA simulations are in purple. From left to right we vary $A_{\rm SN1}$, $A_{\rm SN2}$, $A_{\rm AGN1}$, and $A_{\rm AGN2}$, where the fiducial model is indicated with a solid line, and the minimum and maximum values of the parameter space sampled in CAMELS (see Table \ref{tab:subgrid} for the lower and upper bounds for each parameter) are indicated with a dashed line and a dotted line, respectively. 68th percentiles are shown as the shaded regions around the running medians. Of the four parameters, $A_{\rm SN1}$ is the most efficient at modulating cumulative feedback energy and $f_{\rm CGM}$.}
    \label{fig:efb_fcgm}
\end{figure*}

\begin{figure*}
    \centering
    \includegraphics[width=\linewidth]{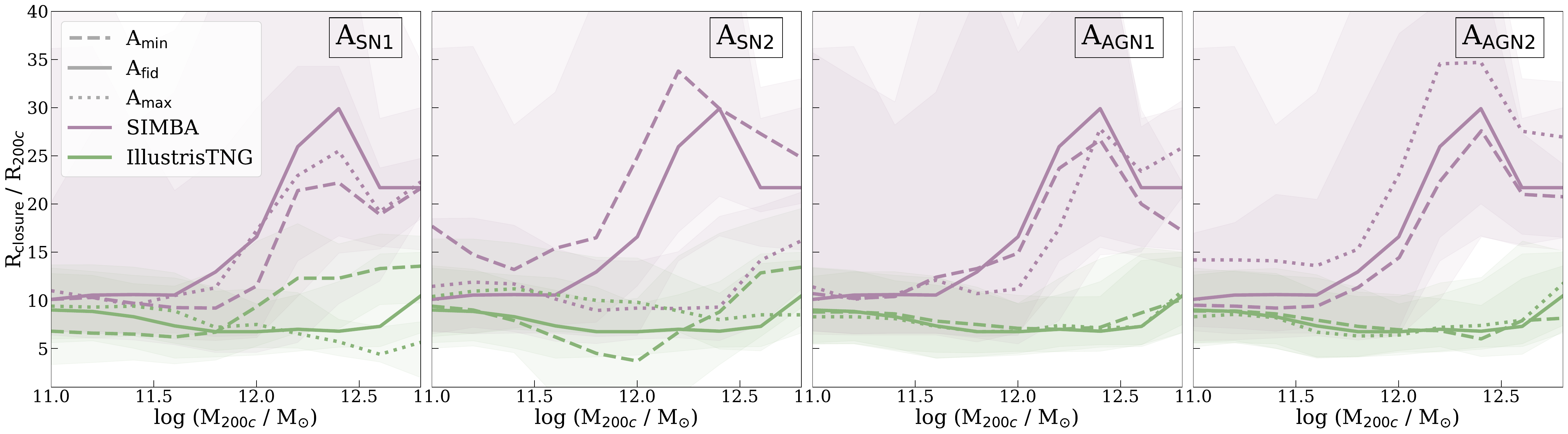}
    \caption{Closure radius (normalized by $R_{200c}$) as a function of halo mass at $z = 0$, shown as running medians. IllustrisTNG simulations are in green and SIMBA simulations are in purple. The fiducial model is indicated with a solid line, the low parameter variation with a dashed line, and the high parameter variation with a dotted line. From left to right we vary $A_{\rm SN1}$, $A_{\rm SN2}$, $A_{\rm AGN1}$, and $A_{\rm AGN2}$. 68th percentiles are shown as the shaded regions around the running medians. The closure radius for SIMBA runs is significantly greater than that of IllustrisTNG runs, due to the more efficient depletion of baryons from the CGM in SIMBA.}
    \label{fig:closure}
\end{figure*}

The total feedback energy is calculated in Section~\ref{sec:data}; here we take the total feedback energy to be the sum of the AGN and SNe feedback modes shown in Figure~\ref{fig:asn1_decompose}, normalized by the binding energy to account for halo mass. This quantity is useful for assessing the impact of parameter variations on the energy being injected into a host halo. In addition to calculating the energetics of the feedback in these simulations, it is also useful to quantify the distribution of baryons as a result of this feedback. The CGM gas fractions are useful in combination with this to identify the trends that we see in the distribution of baryons within the halos, and we plot these quantities as running medians against the halo mass in Figure~\ref{fig:efb_fcgm}. A depleted CGM may indicate that baryons have been expelled from a host halo or that gas inflow is prevented due to strong feedback effects, while a high CGM gas fraction can indicate that outflows are recycled within a halo. The closure radius is another quantity that is useful in this case as the radius within which the baryon fraction is equal to the cosmic baryon fraction. We plot the running medians of the closure radius as a function of halo mass in Figure~\ref{fig:closure}, with a high closure radius indicating a larger spread in baryons and a small closure radius indicating a more concentrated distribution of baryons within the halo. 

Each figure illustrates four columns, depicting parameter changes in $A_\text{SN1}$, $A_\text{AGN1}$, $A_\text{SN2}$, $A_\text{AGN2}$. Lines are dashed, solid, or dotted to represent low, fiducial, and high parameter values, respectively. We plot the running medians with shaded areas for the 16th and 84th percentiles. SIMBA values are indicated in purple, while IllustrisTNG values are in green.

Comparing the two subgrid models generally, we can see that IllustrisTNG overall has higher total feedback energy and higher CGM gas fractions, as we saw in Figure~\ref{fig:fiducial_simba_tng} for fiducial models, but for some parameter variations, these fall below the medians calculated for SIMBA. We note that our calculated values of $f_{\rm CGM}$ are roughly consistent with the mean $f_{\rm b, halo}$ calculated by \cite{Delgado_2023}. The closure radii that we calculate for halos in SIMBA are significantly larger than those for halos in IllustrisTNG, at all masses for all parameter variations. This is especially pronounced at $M_h \gtrsim 10^{12} M_{\odot}$, where there is a sharp peak in the closure radius where the AGN feedback becomes dominant, in agreement with \cite{aryomlou2023}, although with slightly higher values, possibly due to the effects of the CAMELS box sizes or other differences in CAMELS-specific implementations. We also note that the scatter in the closure radius in SIMBA halos is much higher than in IllustrisTNG.

\subsubsection{Stellar feedback effects}

In this section, we discuss the effects of varying the parameters related to galactic winds from stellar feedback on total cumulative feedback, CGM mass fraction, and the closure radius. These parameters are, namely, $A_\text{SN1}$, which controls the mass loading in SIMBA and the energy per unit SFR in IllustrisTNG, and $A_\text{SN2}$, which controls the wind speed in both SIMBA and IllustrisTNG. 

In IllustrisTNG, modulating SNe wind energy per unit SFR factor only leads to slight effects on the upper mass range in terms of total SNe energy but dramatically modulates both thermal and kinetic AGN modes as stellar feedback inhibits the growth of black holes, as we note in Figure~\ref{fig:asn1_decompose}. $A_\text{SN1}$ thus has a pronounced effect on the IllustrisTNG simulations. At low values of this parameter, we see a higher total feedback energy at all masses and a depleted CGM at high halo masses, in addition to an increase in closure radius. This is likely due to the large impact on the thermal AGN. As this parameter is tuned to a lower value, the energy injected by the thermal AGN feedback mode increases, which could lead to material being ejected beyond the CGM and a larger closure radius. 

The reverse behavior of $f_{\rm CGM}$ increasing for low mass halos as $A_{SN1}$ is lowered is a different effect.  Here we are in the regime that AGN kinetic feedback is not contributing at all for these low-mass galaxies, and while thermal feedback is operating and putting significant energy into the CGM (cf. upper left panel of Figure~\ref{fig:asn1_decompose}), this energy is quickly radiating away and matters little for $f_{\rm CGM}$.  We argue that it is the SNe energy that is capable of unbinding gas and altering $f_{\rm CGM}$, and the upper left panel of Figure~\ref{fig:asn1_decompose} shows a surprisingly invariant feedback energy despite $A_{\rm SN1}$ changing by a factor of 16.  If the wind energy per launch event is constant, this would imply that the total mass launched would be constant, and one would expect this to mean the fraction of winds leaving the CGM would also be constant.  However, this is not the case, and a more complex explanation is needed.  

At dwarf masses, we are in the regime where mass loading, $\eta_w \gg 1$.  We turn to an equilibrium model where the SFR plus outflow rate equals the gas inflow rate \citep{Dave2012}.  In the limit of large $\eta$, the outflow rate is nearly the inflow rate, since the SFR is comparatively small.  Therefore, it is not surprising to see relatively constant $E_{\rm FB}$ despite $A_{\rm SN1}$ changing by a factor of 16 and stellar mass also changing by a similar factor, which appears to be the case for the stellar mass fraction for these simulations shown in Figure~14 of \citet{camels_data_release2}.  However, the metallicity in such an equilibrium model is also changing \citep{Finlator2008}, such that at lower $A_{\rm SN1}$, more stars are formed and the metallicity is higher. This metal enrichment leads to more cooling, such that cooling is balancing more significantly the feedback energy that is injected into the CGM (which is independent of $A_{\rm SN1}$ as we show in Figure \ref{fig:asn1_decompose}). Hence, we have a putative explanation for higher $f_{\rm CGM}$ at the lower $A_{\rm SN1}$. Less mass-loaded winds lead to higher stellar masses {\it and} metallicities, which leads to more cooling and thus a greater baryon fraction in the CGM.

The highest CGM gas fraction that we observe across all parameter variations is for the high value of $A_\text{SN1}$ in the IllustrisTNG simulations. This further indicates that this parameter, by inhibiting the growth of black holes at the high-mass end, leads to a much smaller spread in baryons, as also noted in the trend with closure radius. SNe wind energy per unit SFR is the most effective parameter in controlling black hole growth at halo masses above $\sim10^{11.3}M_\odot$, thus leading to a less depleted CGM gas fraction. With high wind energy and at high enough masses, the CGM is enriched with material from galactic winds due to stellar feedback and is not available to accrete onto SMBHs. 

For SIMBA, $A_\text{SN1}$ controls the mass loading factor involved in stellar winds. We thus see a decrease in total feedback energy at low values of this parameter, in addition to lower CGM fractions and closure radii. This may be due to weaker outflows not having enough energy to reach the CGM, thus recycling at smaller radii within the ISM. The most pronounced effect from this is at low masses, where stellar feedback is dominant. 

We see an overall decrease in normalized feedback energy with an increase in $A_\text{SN2}$. Rhe CGM gas fraction for halos at the upper mass end increases with increasing $A_\text{SN2}$. The impact of this parameter variation is much more pronounced in SIMBA halos.

For IllustrisTNG, varying $A_{\rm SN2}$ does not significantly affect the feedback energy; modulation of the wind speed from stellar feedback has a smaller effect on AGN feedback than modulation of wind energy. However, at low masses, the CGM gas fraction is highly affected by this parameter: a higher value leads to a more depleted CGM, which may be due to galactic winds escaping the host halo into the IGM if wind speeds are high enough. At the high mass end, this trend reverses: while all parameter variations result in a relatively low CGM gas fraction, the lowest value of $A_{\rm SN2}$ results in the lowest CGM gas fraction; this could be the result of the wind speeds being low enough to not reach the CGM, rather recycling within the ISM. For a higher value of this parameter, this leads to a flattening of the relation between the CGM gas fraction and the halo mass. 

For SIMBA, we see that a high value of the parameter $A_{\rm SN2}$ leads to a flattened relation between the normalized feedback energy and the halo mass, with a higher feedback energy at the low-mass end and a lower feedback energy at the high-mass end. Likely, as the low-mass end is dominated by stellar feedback, the increase in wind speed translates directly to an increase in energy injection; at the high-mass end, where AGN feedback is dominant, an increase in wind speed prevents material from accreting onto the central SMBH and thus decreases the energy available from this feedback mode. We thus see an increase in the CGM gas fraction at the high-mass end, as less material escapes the host halo from powerful AGN jets. This is consistent with previous trends identified in SIMBA \citep{tillman_exploration_2023-1, Gebhardt_2024}.

Increasing $A_{\rm SN2}$ indeed results in a reduced overall closure radius, as baryon spread is stunted by higher wind speeds. Conversely, a low $A_{\rm SN2}$ value leads to greater feedback energy and reduces the CGM gas fraction for all masses. This is likely because lower wind speeds inhibit materials from leaving the ISM, allowing it to recycle more rapidly. Consequently, the CGM becomes depleted, yet more energy is injected within the host halo's environment.

\subsubsection{AGN feedback effects}

We now discuss the effects of varying parameters related to AGN feedback. $A_\text{AGN1}$ applies to both QSO and jet mode feedback in SIMBA, specifically momentum flux, while in IllustrisTNG it controls kinetic mode feedback, specifically energy per unit of BH accretion rate, or the amount of energy per SMBH accretion event. $A_\text{AGN2}$ controls the jet mode feedback only in SIMBA and affects the jet speed, and in IllustrisTNG this parameter controls the ejection speed and burstiness of the kinetic mode feedback.

Increasing $A_\text{AGN1}$ boosts the efficiency with which a black hole converts mass into energy. However, this has minimal impact on the total feedback injection in IllustrisTNG; perhaps due to self-regulation of the black hole. \cite{camels_data_release2} show that there is a slight suppression of high mass ($M_{\rm BH} > 10^8 M_{\odot}$) black holes with increased $A_\text{AGN1}$. Modulating $A_\text{AGN1}$ does, however, strongly affect the CGM gas fraction at the high halo mass end. A lower value of $A_\text{AGN1}$ leads to a more baryon-rich CGM. The lower energy of the AGN feedback events means that the material does not escape beyond the CGM.

In SIMBA, $A_\text{AGN1}$ affects both radiative/QSO and jet modes of AGN feedback. As a prefactor of the momentum flux and scaling with the outflow velocity, this parameter changes the amount of material ejected from AGN feedback, with a larger effect on the radiative mode. We see that with a lower value or lower momentum flux, the total feedback energy is highest, and the CGM gas fraction is lowest. With a high value, the total feedback energy decreases and the peak in the CGM gas fraction shifts to higher masses. 

In IllustrisTNG, $A_\text{AGN2}$ controls the burstiness and strength of the AGN feedback. Increasing $A_\text{AGN2}$ decreases the frequency of kinetic feedback events but makes them individually stronger. Consequently, Figure~\ref{fig:efb_fcgm} illustrates that changing $A_\text{AGN2}$ minimally affects the overall cumulative feedback energy. However, for halos with mass $M_h \gtrsim 10^{12} M_{\odot}$, where the kinetic feedback mode is active, there is a notable impact on the CGM gas fraction. More sporadic feedback, defined by larger $A_\text{AGN2}$ values, results in a more depleted CGM, as each event has greater energy to expel baryons from the halos, despite their reduced frequency.

In SIMBA, $A_\text{AGN2}$ affects the jet speed of the jet mode AGN feedback. Varying this parameter does not significantly change the total feedback energy, possibly due to the small box sizes implemented in CAMELS inhibiting the formation of the systems that would be most strongly impacted by this parameter. However, a higher value leads to a more depleted CGM in all masses, possibly by ejecting material beyond the CGM; this agrees with the results of the last panel in Figure~\ref{fig:closure}: we see a larger closure radius with faster AGN jet speeds at all masses. With an increase in jet outflow velocity, the ejected material escapes the host halo, thus depleting the CGM and increasing the radius in which the cosmic baryon fraction is contained. The trend in variation is constant across all masses for this parameter for the CGM gas fraction; a higher value leads to a more depleted CGM, and a lower value leads to a less depleted CGM. As AGN feedback is not dominant at low-masses (and AGN activity is not present at the low-mass end), this suggests a couple of possibilities. First, that the AGN mode of feedback is complexly interacting with stellar feedback, with higher jet speeds setting off a series of effects possibly leading to more efficient stellar feedback. Alternatively, faster AGN jets from massive halos could potentially affect low-mass halos due to environmental effects.

Overall, varying $A_\text{AGN1}$ and $A_\text{AGN2}$ has a much more dramatic effect in SIMBA than in IllustrisTNG, where it seems to have little effect on the feedback energetics. In SIMBA, there is a significant effect on the CGM gas fraction at all halo masses in SIMBA. However, while the variation appears to have little effect on the energetics in IllustrisTNG, there is a significant effect on the CGM gas fraction in halos with mass above $\sim 10^{12} M_{\odot}$, with an increase in the parameter values (increase in feedback efficiency) leading to a lower CGM gas fraction or a more effective baryon depletion. Additionally, for IllustrisTNG other AGN parameters in the extended 1P-28 set have a more significant effect on both the feedback energetics and CGM gas fraction (see \ref{sec:appendix} for details).

\subsection{Redshift evolution}

\begin{figure*}
    \centering
    \includegraphics[width=0.95\linewidth]{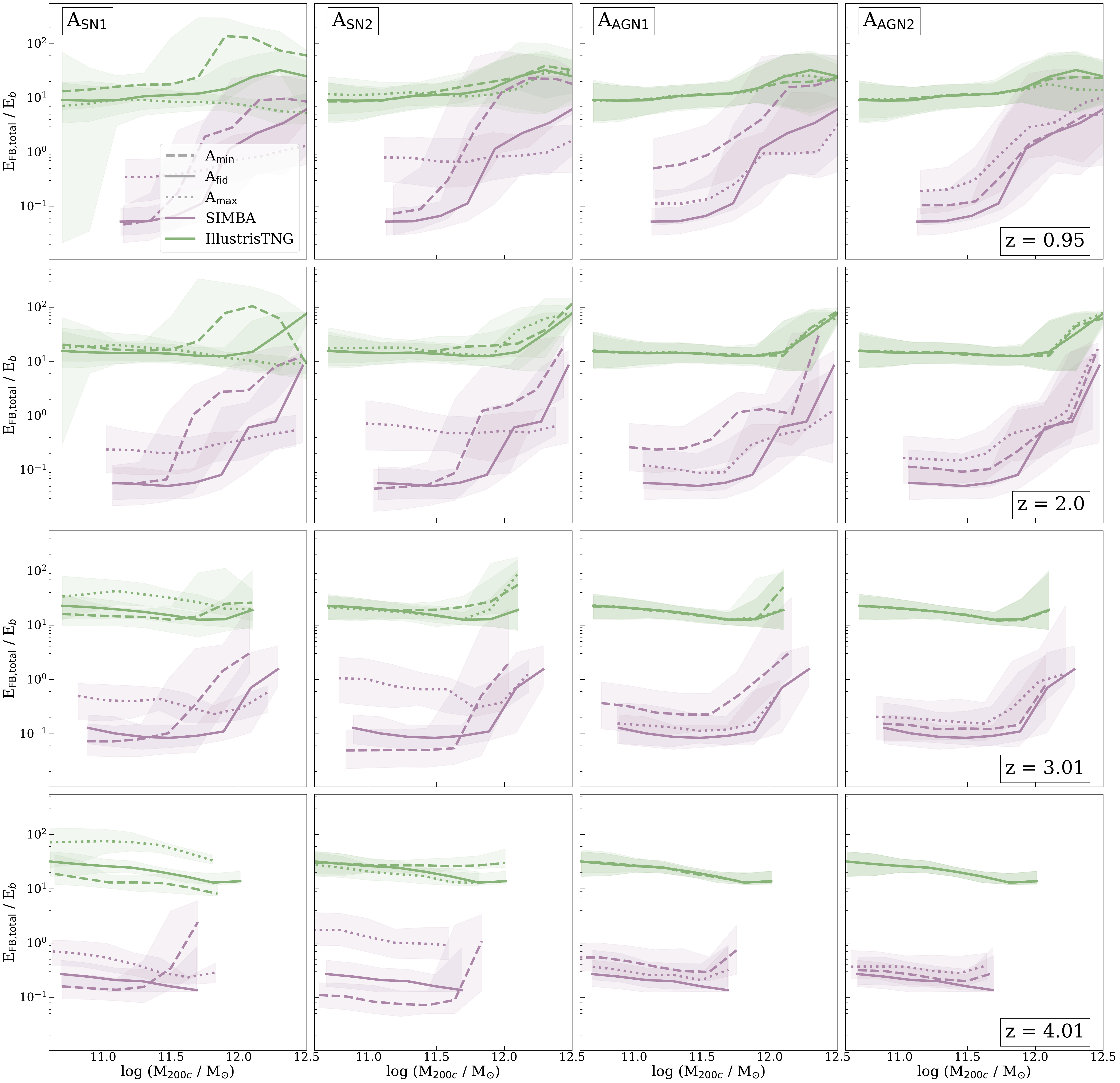}

    \caption{Same as the top row of Figure~\ref{fig:efb_fcgm}, for $z=0.95$ (top), $z=2.0$, $z=3.01$, and $z=4.01$ (bottom), showing cumulative feedback energy (normalized by halo binding energy) as a function of halo mass, shown as running medians. IllustrisTNG simulations are in green and SIMBA simulations are in purple. The fiducial model is indicated with a solid line, the low parameter variation with a dashed line, and the high parameter variation with a dotted line. From left to right we vary $A_{\rm SN1}$, $A_{\rm SN2}$, $A_{\rm AGN1}$, and $A_{\rm AGN2}$. 68th percentiles are shown as the shaded regions around the running medians. At low redshift ($\sim 2$), with the onset of significant AGN feedback, baryon dynamics become more complex and interlaced.}
    \label{fig:efb_z}
\end{figure*}

\begin{figure*}
    \centering
    \includegraphics[width=0.95\linewidth]{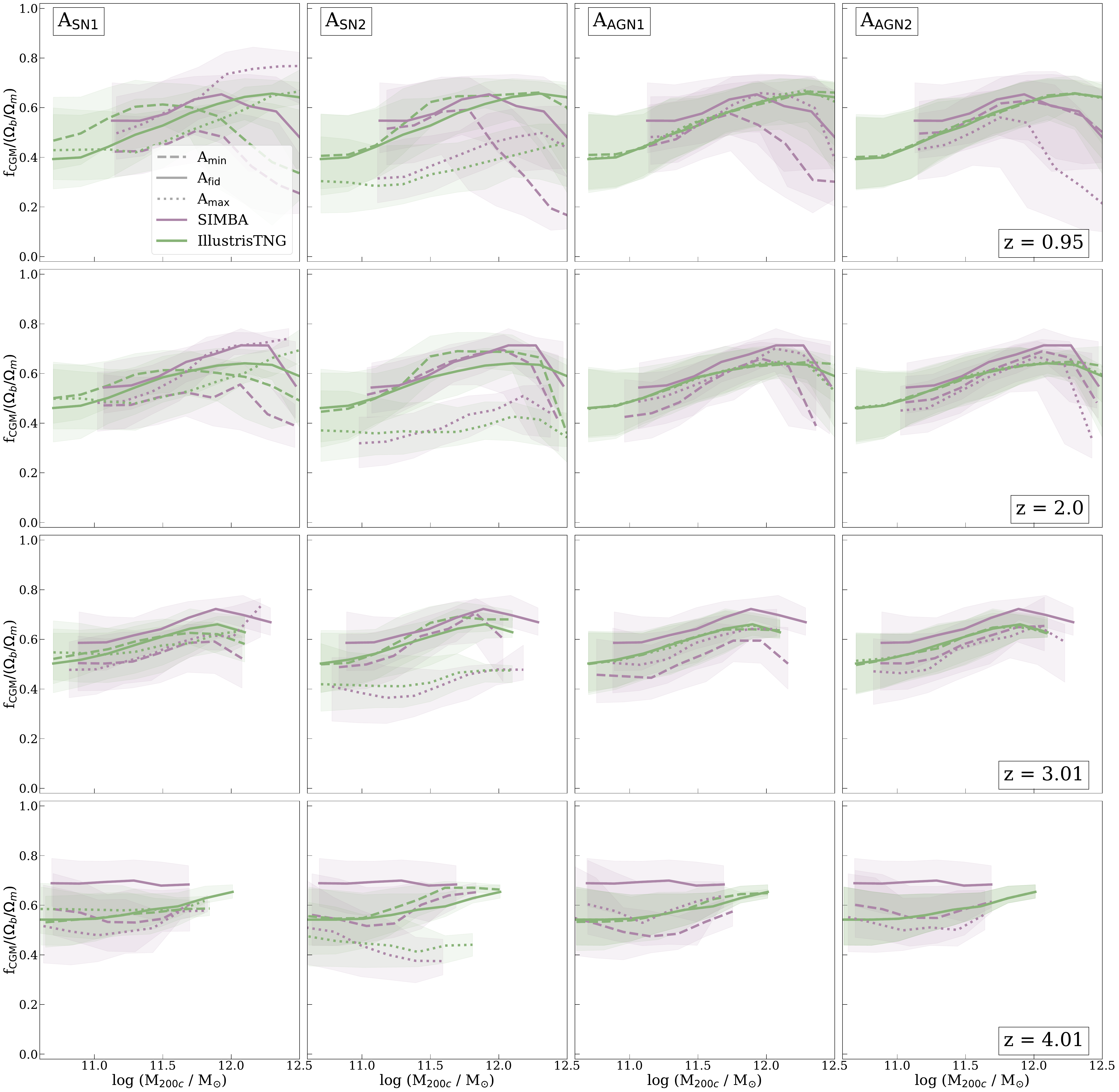}

    \caption{Same as the bottom row of Figure~\ref{fig:efb_fcgm}, for $z=0.95$ (top), $z=2.0$, $z=3.01$, and $z=4.01$ (bottom), showing the CGM gas mass fraction $f_{\rm CGM}$ (normalized by $f_{\rm b, cosmic} = \Omega_b/\Omega_m$) as a function of halo mass, shown as running medians. IllustrisTNG simulations are in green and SIMBA simulations are in purple. The fiducial model is indicated with a solid line, the low parameter variation with a dashed line, and the high parameter variation with a dotted line. From left to right we vary $A_{\rm SN1}$, $A_{\rm SN2}$, $A_{\rm AGN1}$, and $A_{\rm AGN2}$. 68th percentiles are shown as the shaded regions around the running medians. In both SIMBA and IllustrisTNG, modulation of AGN feedback parameters has the most dramatic effect on $f_{\rm CGM}$ at all redshifts.}
    \label{fig:fcgm_z}
\end{figure*}

Until this point, our focus has been on the energetics of feedback and their influence on the halo properties at $z = 0$. We now extend our analysis to higher redshifts, focusing on four additional redshift snapshots at $z = 1, 2, 3, \text{and } 4$. We are particularly interested in $z = 2$ as this is where the cosmic SFR and AGN activity peak \citep{madau_2014}. Figures \ref{fig:efb_z} and \ref{fig:fcgm_z} show the same comparisons between the simulations for cumulative injected feedback energy and CGM gas fractions with the four parameter variations as Figure \ref{fig:efb_fcgm}, extended to higher redshifts (from top to bottom, $z=0.95$, $z=2.0$, $z=3.01$, and $z=4.01$). We extend the mass cut for the halos in the extended redshift analysis to the 500 most massive halos and perform the calculations described in Section~\ref{sec:data}. 

First, we note that, across all redshifts, IllustrisTNG maintains a greater cumulative feedback energy, as we noted in the fiducial comparisons at $z = 0$ in Figure~\ref{fig:fiducial_simba_tng}. This difference becomes especially pronounced in halos with a mass below $\sim 10^{11.7} - 10^{12} M_{\odot}$, depending on $z$ and the parameter values, where the energy values are two orders of magnitude higher than for SIMBA. Thus, at high redshift, where halos above this mass range have not yet formed in the limited CAMELS volume, by eye the disparity appears drastic. However, the difference is not as dramatic for the CGM mass fraction. This is due to the thermal AGN mode that grows black holes below $10^8\;M_{\odot}$ in IllustrisTNG being inefficient at ejecting baryons out of the halo. For the most part, as we noted at $z = 0$ SIMBA has a more depleted CGM. 

At redshift $z = 4$, both SIMBA and IllustrisTNG predominantly exhibit the effects of SNe feedback, which generally decreases slightly with increasing mass. In SIMBA, there is a slight rise in feedback energy at the high-mass end, indicating the onset of AGN feedback. Specifically, with adjustments to $A_{\rm SN1}$, we observe a significant peak of AGN feedback even at $z = 4$. In SIMBA, the jet mode of AGN feedback activates earlier than in IllustrisTNG due to a lower black hole mass threshold for initiating the kinetic mode ($M_{\rm BH, thresh} = 10^{7.5} M_{\odot}$ in SIMBA, compared to $M_{\rm BH, thresh} = 10^{8} M_{\odot}$ in TNG). Indeed, ny the time we reach $z = 3$, IllustrisTNG also begins to show enhanced feedback energy at the high-mass end.

In IllustrisTNG, unlike for $z = 0$, when modulating $A_{\rm SN1}$ to higher values, we see an overall increase in the feedback energy in $z = 4$ and $z = 3$ in Figure~\ref{fig:efb_z}. It is not until $z = 2$ that we see that the flip begins to occur where increasing $A_{\rm SN1}$ leads to a decrease in overall feedback at $M_{h} \gtrsim 10^{11.6} M_{\odot}$. This corresponds to the time when in the CAMELS volume large enough halos have formed to host AGN that turn on the kinetic mode of feedback. Correspondingly, for the low value of $A_{\rm SN1}$, we see the first true peak in the feedback energy due to AGN, limiting supernova feedback allowing black holes to grow and evolve faster. Now looking at the CGM gas fractions in Figure~\ref{fig:fcgm_z} correspondingly $f_{\rm CGM}$ is greater for the lower feedback runs of IllustrisTNG.

In SIMBA, \cite{sorini_simba_2020} find that at $z \sim 2-3$ it is stellar feedback that is the primary physical driver in determining the properties of the CGM around quasars. Indeed, we see that in SIMBA the modulation of $A_{\rm SN2}$ in particular, and to a lesser extent $A_{\rm SN1}$ has a more dramatic impact on $f_{\rm CGM}$ than the modulation of AGN feedback parameters.

We can further contextualize the results at $z = 2$ by considering again the \cite{wright2024} outflow and inflow results for the original IllustrisTNG and SIMBA simulations. They note that for IllustrisTNG outflows are stalled from the ISM scale to the CGM scale at $z = 2$, which explains that while there is a high cumulative feedback in IllustrisTNG, the CGM gas mass fraction is greater than that of SIMBA. This is due to the velocities imparted to the wind particles at injection, which are almost always below the halo escape velocity, which confined them to stay within the halo. In SIMBA, the outflow rate actually increases with scale for halos with $10^{11.5} M_{\odot} < M_{h}  < 10^{13} M_{\odot}$, indicating mass entrainment. In general, we see that at $z = 2$, SIMBA is more efficient in pushing baryons despite having lower cumulative feedback energy, just as at $z = 0$.

To summarize, our investigation into the redshift evolution of feedback energetics and $f_{\rm CGM}$ reveals that modulating SNe feedback generally inhibits black hole growth and delays the onset of AGN feedback modes. Before the onset of AGN feedback, the energetics  scale with the wind strength of SNe and modulate $f_{\rm CGM}$ accordingly. With the onset of AGN feedback, dynamics become more complex and interlaced; adjusting a parameter might not produce the anticipated outcome.

\section{Discussion} \label{sec:disc}
Differences between SIMBA and IllustrisTNG are inevitable: with different feedback prescriptions, methods of calculation for the different available parameters, and different halos samples, our aim is to focus on general trends in the data and the effects of varying the feedback parameters in this paper. We also note that we do not effectively capture every component of energy injected by feedback effects; however, the focus of this paper is the qualitative differences between the simulations and when varying astrophysical parameters related to feedback. Additionally, while the feedback processes related to AGN and stellar feedback are not able to be fully quantitatively separated, by modulating the astrophysical parameters, we can try to disentangle the full effects of these processes. Rather than examining the coupling between the parameters directly, we change each individually to distinguish effects from each.

Another caveat that we note is that we measure only the energy injected from outflowing gas due to feedback and do not track inflowing gas, another contributor to CGM gas fractions. For an in-depth discussion of the effects of inflowing gas \citep[see e.g.,][]{wright2024}.

Finally, we note that the box size used in CAMELS is smaller than in the original IllustrisTNG and SIMBA simulations. This restricted box size limits the emergence of larger mass systems on the cluster scales. Consequently, this constraint prevents us from fully observing the influence of parameters such as $A_\text{AGN2}$ on scales beyond the highest mass thresholds considered in this study, specifically $M_{200c} \sim 10^{12.8}M_{\odot}$. A potential approach to extend our analysis to higher mass halos is through the use of the zoom-in simulations with parameter variations presented by \cite{Lee_2024}.

While it is outside the scope of this paper, future work can build on our findings and work towards a better understanding of how the efficiency of stellar feedback leads to the effects that we see in AGN feedback, and vice versa, through a full exploration of the coupling between these four astrophysical parameters. 

\section{Conclusions}\label{sec:conclusion}

We conduct a thorough investigation of the baryonic feedback energetics in the CAMELS-SIMBA and CAMELS-IllustrisTNG simulation suites. We examine how the subgrid implementation differences between the two suites affect the pure cumulative feedback energetics as well as the baryon distribution around halos, using the CGM gas fraction and the closure radius as key metrics. In addition, we systemically investigate how varying key parameters within CAMELS-SIMBA and CAMELS-IllustrisTNG affect each of these quantities. We summarize our findings as follows:

\begin{itemize}
    \item Comparing SIMBA and IllustrisTNG we see that, overall, IllustrisTNG displays greater cumulative feedback energy over all the halos (Figures~\ref{fig:fiducial_simba_tng}~and~\ref{fig:efb_fcgm}). However, the closure radii of SIMBA halos are much larger, indicating that baryons are being pushed further out from the halos (Figure~\ref{fig:closure}). While feedback overall is not stronger, it more effectively couples to baryons and is more efficient at distributing baryons within and beyond the host halo; this is possibly due to jet mode and bipolar outflows and ejections extending out into the IGM, while IllustrisTNG feedback modes, which are isotropic, may be more confined to the host halo.
    
    \item Of the four astrophysical parameters, SNe wind energy per unit SFR (parameter $A_\text{SN1}$) is the most effective parameter in inhibiting black hole growth at halo masses above $\sim10^{11.3}M_\odot$ in IllustrisTNG and thus leading to a less depleted CGM gas fraction (Figures~\ref{fig:asn1_decompose}~and~\ref{fig:efb_fcgm}). In SIMBA, increasing $A_\text{SN1}$ (which controls the mass loading factor of SNe) has a similar effect on the growth of black holes at halo masses below $\sim10^{11.0}M_\odot$. This parameter controlling the efficiency of stellar feedback leads to the suppression of black hole mass growth and thus to a delay in the onset of AGN feedback to activate in these high-mass ($M_h \gtrsim 10^{11.7} M_{\odot}$) halos in both simulations.
            
    \item More efficient stellar feedback (e.g., higher $A_\text{SN1}$ parameter) is correlated with weaker AGN feedback in IllustrisTNG, due to the suppression of black hole growth. On the other hand, in SIMBA, despite the suppression of the number of low-mass black holes, more efficient stellar feedback is correlated with an enhanced number of high-mass ($M_{\rm BH} > 10^{8} M_{\odot}$) black holes as well as stronger AGN feedback (Figure~\ref{fig:asn1_decompose}). 

    \item Varying the $A_{\rm AGN1}$ and $A_{\rm AGN2}$ feedback parameters has little impact on the feedback energetics in IllustrisTNG. However, a slight change is seen in the CGM gas fraction at large halo masses, with a decrease in the parameter values (decrease in feedback efficiency) leading to a higher CGM gas fraction. Additionally, a selection of AGN parameters from the extended 1P 28 parameter set have significant effects on the feedback energy and the CGM gas fraction, as discussed in \ref{sec:appendix}. In SIMBA, the $A_{\rm AGN1}$ and $A_{\rm AGN2}$ parameters influence CGM gas fractions across halos of all mass ranges. This suggests that, although the parameters are primarily used for adjusting AGN feedback modes, there are likely complex, non-linear interactions between SNe and AGN feedback modes (Figure~\ref{fig:efb_fcgm}).

    \item  In IllustrisTNG for $z > 2$, the parameter $A_\text{SN1}$ inhibits black hole growth, resulting in AGN feedback being either inactive or negligible. In this regime, SNe feedback is predominant, and directly adjusting $A_\text{SN1}$ enhances total cumulative feedback, since the intricate interplay with AGN feedback is absent. Similarly, in SIMBA at higher redshifts, while AGN feedback is not the main contributor, it is still significant at the highest mass end. Overall, at low redshifts, the emergence of AGN feedback brings about more complex and interconnected dynamics; adjusting one parameter may not yield the anticipated outcomes (Figures~\ref{fig:efb_z}~and~\ref{fig:fcgm_z}).
    
\end{itemize}

These results indicate that AGN and stellar components are intricately connected, and moreover seemingly couple differently in different simulations. Further investigation of this coupling may allow us to more fully understand how these different modes of feedback are connected in galaxies.

\section*{Acknowledgements}
The CAMELS simulations were performed on the supercomputing facilities of the Flatiron Institute, which is supported by the Simons Foundation. This work is supported by the NSF grant AST 2206055 and the Yale Center for Research Computing facilities and staff. IM acknowledges support from the Dean's Emerging Scholars Research Award from the Yale Graduate School of Arts \& Sciences. DAA acknowledges support by NSF grant AST-2108944, NASA grant ATP23-0156, STScI grants JWST-GO-01712.009-A and JWST-AR-04357.001-A, Simons Foundation Award CCA-1018464, and Cottrell Scholar Award CS-CSA-2023-028 by the Research Corporation for Science Advancement. PS acknowledges support from the YCAA Prize Postdoctoral Fellowship. 

\vspace{5mm}

\appendix

\section{CAMELS-TNG 1P-28 Set Parameter Descriptions}

In Table \ref{tab:1P28_desc} we provide a complete description of the 28 parameters included in the extended 1P set of CAMELS-IllustrisTNG \citep{camels_data_release2}. In the first column, we give the name of the parameter with the corresponding symbol in the second column. In the third column, we provide a description of the physical meaning of each parameter and point to where it is implemented in the subgrid model. In the last three columns, we give the fiducial value, the variation range, and specify if it is varied linearly or logarithmically. If the parameter is unitless, no units are given, otherwise, the relevant units are listed in the fiducial value column.

\begin{longtable}{|>{\raggedright}p{2.45cm}|p{1.35cm}|p{7.0cm}|p{1.35cm}|p{2.3cm}|p{1.35cm}|}
 \hline
 \multicolumn{6}{|c|}{CAMELS-TNG 1P-28 Parameter List} \\
 \hline
 Parameter & Symbol & Description & Fiducial & Range & Variation \\
 \hline
 
 Omega0 & $\Omega_{m}$ & The standard cosmological parameter omega matter, the $z=0$ cosmic matter density in units of the critical density & 0.3 &  [0.1, 0.5] & linear \\
 
 Sigma8 & $\sigma_8$ & The standard cosmological parameter sigma 8, the RMS of the $z=0$ linear overdensity in spheres of radius $8 h^{-1}$Mpc & 0.8  & [0.6, 1.0] & linear \\

 Wind Energy In 1e51 erg & $A_{\rm SN1}$ & Normalization factor for the energy in galactic winds per unit star-formation, denoted as a prefactor to $\Bar{e}_w$ in Eq.~3 in \citet{pillepich_illustris_sim_2018}. & 1 & [0.25, 4] & log \\

 Variable Wind Velocity Factor & $A_{\rm SN2}$ & Normalization factor for the galactic wind speed, denoted as a prefactor to $\kappa_w$ in Eq.~1 of \citet{pillepich_illustris_sim_2018}. & 1 & [0.5, 2] & log \\

 Radio Feedback Factor & $A_{\rm AGN1}$ & Normalization factor for the energy in AGN feedback, per unit accretion rate, in the low-accretion state. Implemented as a pre-factor in front of the RHS of Eq.~8 in \citet{Weinberger_2017}. & 1 & [0.25, 4.0] & log \\

 Radio Feedback Reorientation Factor & $A_{\rm AGN2}$ & Normalization factor for the frequency of AGN feedback energy release events in the low-accretion state, denoted a prefactor to $f_{\rm re}$ in Eq.~13 of \citet{Weinberger_2017}. & 1 & [0.5, 2] & log \\
  
 Omega Baryon & $\Omega_b$ & The standard cosmological parameter $\Omega_b$, the $z=0$ cosmic baryon density in units of the critical density. & 0.049 & [0.029, 0.069] & linear \\

 Hubble Parameter & $H_0$ & The standard Hubble constant, in units of 100 km s\textsuperscript{-1} Mpc\textsuperscript{-1}. & 0.6711 & [0.4711, 0.8711] & linear \\
 
 Name & $n_s$ & The standard cosmological parameter $n_s$, the spectral index of the initial fluctuations, described by the power spectrum. & 0.9624 & [0.7624, 1.1624] & linear \\

 Max SFR Timescale & $t^*_0$ & The timescale for star-formation at the density threshold of star-formation, denoted as $t^*_0$ in \citet{Springel_Hernquist_2003} & 2.27 Gyr & [1.135, 4.54] & log \\

 Factor For Softer EQS & $q_{\rm EOS}$ & An interpolation factor between the effective equation of state for the star-forming gas obtained from the \citet{Springel_Hernquist_2003} subgrid model and an isothermal equation of state with $10^4K$, denoted as $q_{\rm EOS}$ in \citet{springel_modelling_2005} & 0.3 & [0.1, 0.9] & log \\

 IMF Slope & n/a & The power law index of the stellar initial mass function above $1 M_{\odot}$. Below that mass, the IMF is kept fixed following \citet{Chabrier_2003}. & -2.3 & [-2.8, -1.8] & linear \\

 SNII Min Mass Msun & $M_{\rm SNII,min}$ & The lower threshold for the mass of a star that produces a supernova explosion, denoted as $M_{\rm SNII,min}$ in \citet{Vogelsberger_2013}. & 8 $M_{\odot}$ & [4, 12] & linear \\

 Thermal Wind Fraction & $\tau_w$ & The fraction of the galactic wind feedback energy that is injected thermally, denoted as $\tau_w$ in \citet{pillepich_illustris_sim_2018}. & 0.1 & [0.025, 0.4] & log \\

 Variable Wind Spec Momentum & $\rm mom_w$ & A normalization factor for the specific momentum in galactic winds per unit star formation, denoted as $\rm mom_w$ in \citet{Vogelsberger_2013}. & 0~km/s  & [0, 4000] & linear \\

 Wind Free Travel Dens Factor & n/a & Sets the gas density around collisionless galactic wind particles at which they recouple back into the hydrodynamics. & 0.05 & [0.005, 0.5] & log \\

 Min Wind Vel & $v_{\rm w,min}$ & The minimum value imposed for the galactic wind speed, denoted as $v_{\rm w,min}$ in \citet{pillepich_illustris_sim_2018}. & 350 km/s & [150, 550] & linear \\

 Wind Energy Reduction Factor & $f_{\rm w,Z}$ & Normalization factor for the energy of galactic winds at high metallicity compared to low metallicity, denoted by $f_{\rm w,Z}$ in \citet{pillepich_illustris_sim_2018}. & 0.25 & [0.0625, 1.0] & log \\
 
 Wind Energy Reduction Metallicity & $Z_{\rm w,ref}$ & Sets the metallicity at which the transition from high to low energy galactic winds occurs, dentoed as $Z_{\rm w,ref}$ in \citet{pillepich_illustris_sim_2018}. & 0.002 & [0.0005, 0.008] & log \\

 Wind Energy Reduction Exponent & $\gamma_{w,Z}$ & Controls the abruptness in metallicity of the transition between high- and low-energy galactic winds occurs, denoted by $\gamma_{w,Z}$ in \citet{pillepich_illustris_sim_2018}. & 2 & [1, 3] & linear\\

 Wind Dump Factor & $1 - \gamma_w$ & The fraction of metals in a star-forming cell that are ejected into a galactic wind that are deposited in neighboring star-forming cells prior to ejection, denoted as $1 - \gamma_w$ in \citet{pillepich_illustris_sim_2018}. & 0.6 & [0.2, 1.0] & linear \\

 Seed Black Hole Mass & $M_{\rm seed}$ & The mass of the seed supermassive black holes, as described in \citet{Vogelsberger_2013}. & $8 \cdot 10^5$ $M_{\odot}$ & $2.5 \cdot$[$10^5$, $10^6$] & log \\

 Black Hole Accretion Factor & n/a & A normalization factor for the Bondi rate for the accretion onto supermassive black holes, implemented as a pre-factor in front of the RHS of Eq.~2 in \citet{Weinberger_2017} & 1.0 & [0.25, 4.0] & log \\

 Black Hole Eddington Factor & n/a & A normalization factor for the limiting Eddington rate for the accretion onto supermassive black holes, implemented as a pre-factor in front of the RHS of Eq.~3 of \citet{Weinberger_2017}. & 1.0 & [0.1, 10] & log\\

 Black Hole Feedback Factor & $\epsilon_{\rm f,high}$ & A normalization factor for the energy in AGN feedback, per unit accretion rate, in the high-accretion state, and is denoted as $\epsilon_{\rm f,high}$ in Eq.~7 of \citet{Weinberger_2017}. & 0.1 & [0.025, 0.4] & log\\ 

 Black Hole Radiative Efficiency & $\epsilon_r$ & The radiative efficiency of AGN feedback, the fraction of the accretion rest-mass that is released in the accretion process. & 0.2 & [0.05, 0.8] & log\\ 

 Quasar Threshold & $\chi_0$ & The Eddington ratio (at the pivot mass of $10^8 M_{\odot}$) that serves as the threshold between the low-accretion and high-accretion states of AGN feedback, denoted by $\chi_0$ in Eq.~5 of \citet{Weinberger_2017} & 0.002 & $6.3 \cdot$$[10^{-5}$, $10^{-2}$] & log \\ 

 Quasar Threshold Power & $\beta$ & The power-law index of the scaling of the low to high accretion state threshold with black hole mass, denoted by $\beta$ in Eq.~5 of \citet{Weinberger_2017}. & 2.0 & [0.0, 4.0] & linear \\ 

 \hline
 \caption{Overview of the 28 parameters that are varied in the CAMELS-TNG 1P-28 set. Columns are: name of parameter, symbol, physical description, fiducial value, range of variation, and sampling in linear or logarithm space.}
    \label{tab:1P28_desc}\\
 
\end{longtable}

\section{CAMELS-TNG 1P-28 Set Feedback Energetics and CGM Gas Fraction} \label{sec:appendix}
In addition to the original 1P set of CAMELS-SIMBA and CAMELS-IllustrisTNG, we analyze the extended CAMELS-IllustrisTNG 1P set, containing 28 parameters, described in detail in the previous section. We calculate the total cumulative feedback energy and the CGM gas fraction. We do not include closure radius calculations due to computational constraints.

\begin{figure*}
    \centering
    \includegraphics[width=\linewidth]{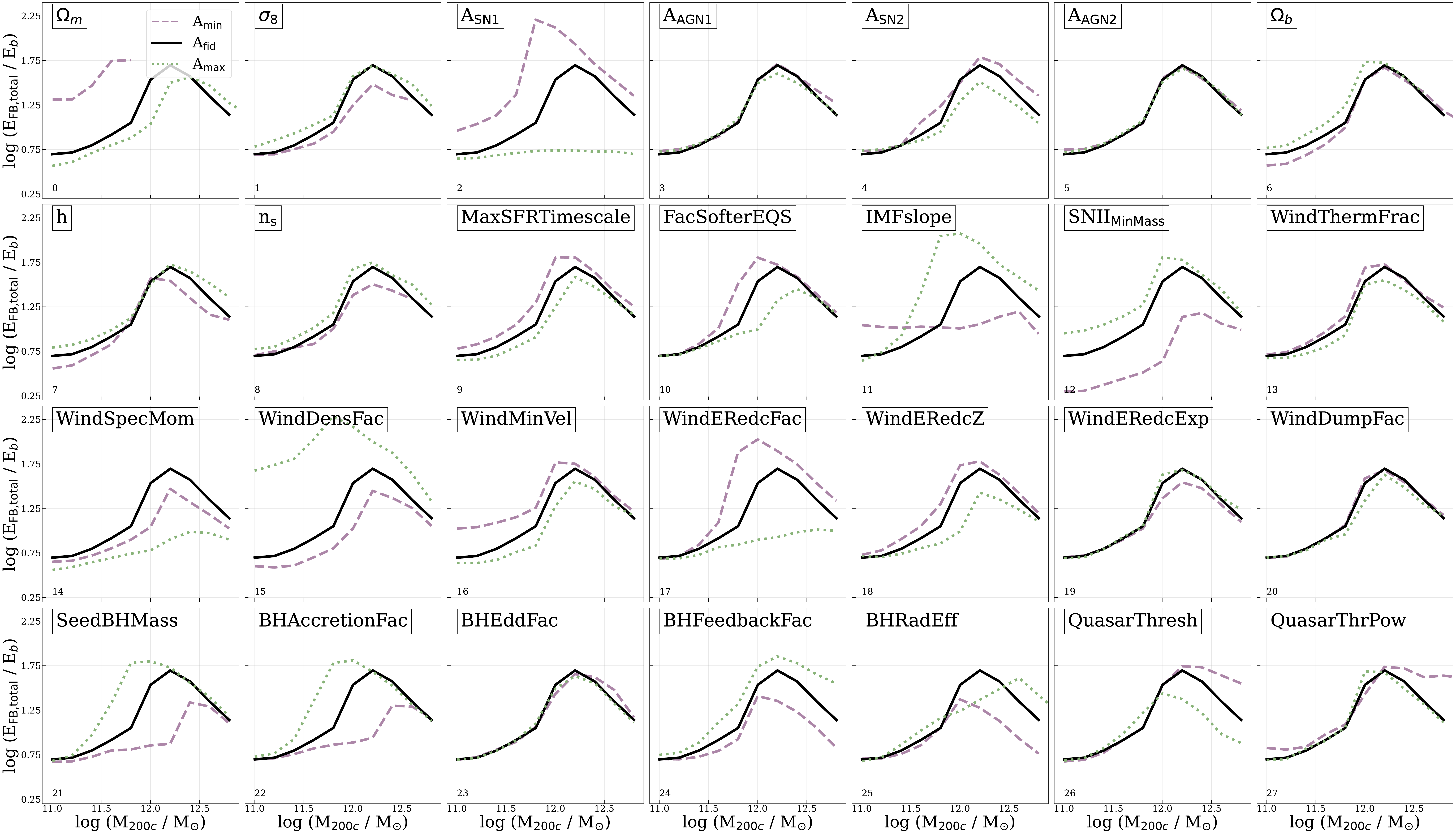}
    \caption{Total cumulative feedback energy (normalized by halo binding energy) as a function of halo mass at $z=0$, shown as running medians, for the 28 parameters of the IllustrisTNG extended 1P set. CGM gas fraction $f_{\rm CGM}$ (normalized by $f_{\rm b, cosmic} = \Omega_b/\Omega_m$) as a function of halo mass at $z=0$, shown as running medians. The fiducial model is indicated with a black line, the low parameter variation with a purple line, and the high parameter variation with a green line. In addition to $A_{rm SN1}$, several other parameters have significant impact on the cumulative feedback energetics, notably shifting the halo mass at which we begin to see the impact of AGN feedback.}
    \label{fig:efb_1p28}
\end{figure*}

\begin{figure*}
    \centering
    \includegraphics[width=\linewidth]{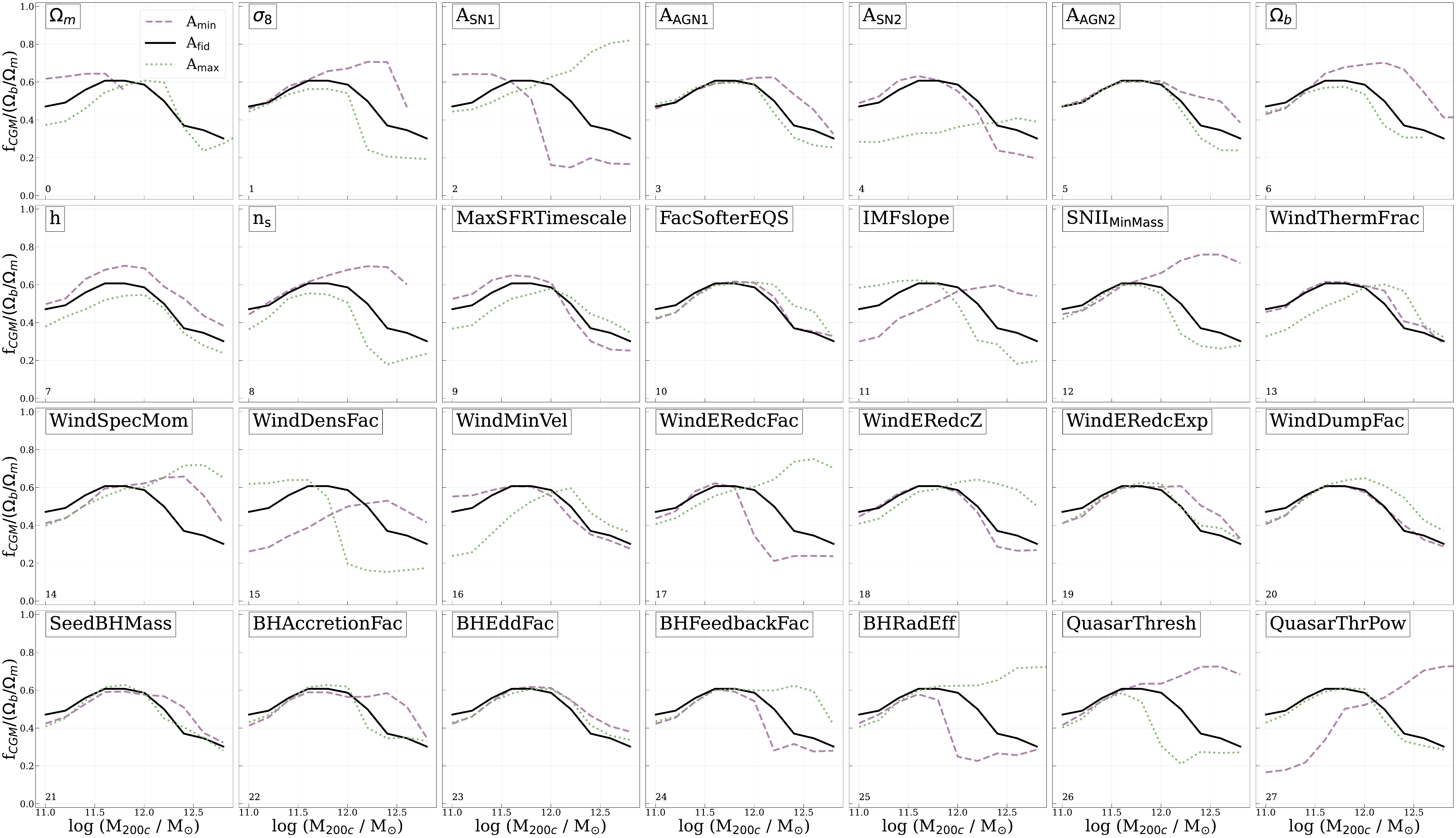}
    \caption{CGM gas fraction $f_{\rm CGM}$ (normalized by $f_{\rm b, cosmic} = \Omega_b/\Omega_m$) as a function of halo mass at $z=0$, shown as running medians, for the 28 parameters of the IllustrisTNG extended 1P set. The fiducial model is indicated with a black line, the low parameter variation with a purple line, and the high parameter variation with a green line. Most parameters preserve the general trend of decreasing $f_{\rm CGM}$ with halo mass, but these trends are reversed for select variations.}
    \label{fig:fcgm_1p28}
\end{figure*}

In Figures~\ref{fig:efb_1p28} and \ref{fig:fcgm_1p28}, we show the total cumulative feedback energy and the CGM gas mass fraction at $z = 0$, respectively as a function of halo mass and modulated by each of the 28 parameters. In these figures, the low-value variation simulation is shown with the purple line, the fiducial model with the black line, and the high-value simulation with the green line.

First, focusing on Figure~\ref{fig:efb_1p28} we note that most parameters have a significant visible impact on the cumulative feedback energy, except for the two original AGN parameters, $A_{\rm AGN1}$ and $A_{\rm AGN2}$ as well as a couple of the new parameters: the Black Hole Eddington Factor and the Wind Dump Factor. Some factors have a notably stronger effect across all halo masses than most others. These are $\Omega_m$, $A_{\rm ASN1}$, the IMF slope, the SNII Min Mass, the Wind Density Factor, the Wind Energy Reduction Exponent, the Seed BH Mass, and the BH Accretion Factor. Of these, one is a cosmological parameter, five modulate SNe feedback, and two are related to AGN feedback. Focusing on parameters modulating AGN feedback, we note that parameters such as Quasar Thresh and Quasar Threshold Power have a significant effect above a threshold mass of about $M_{h} = 10^{12} M_{\odot}$ which makes sense as these are modulating the kinetic feedback mode which only turns on at high enough mass. Two Seed BH Mass and BH Accretion Factor parameters tune where AGN feedback turns on, with higher values pushing that turning point to lower halo mass.

Now, focusing on Figure~\ref{fig:fcgm_1p28}, we note that most parameters have a noticeable impact when varied, even parameters that did not seem to significantly vary total feedback energetics. The most significant parameters here are: $A_{\rm SN1}$, $A_{\rm SN2}$, the IMF slope, the SNII Min Mass, the Wind Density Factor, the Wind Reduction Factor, the BH Radiative Efficiency, the Quasar Threshold, and the Quasar Threshold Power. Six of these parameters relate to the SNe feedback, while three modulate the AGN feedback. Some of these factors, specifically $A_{\rm SN1}$, BH Radiative Efficiency, and Quasar Threshold Power, are especially interesting as they do not follow the usual trend of a decrease in $f_{\rm CGM}$ after halo mass of $\sim 10^{12} M_{\odot}$. For these in particular, $f_{\rm CGM}$ remains constant or continues to increase as a function of mass. For other parameters, this effect is less dramatic, and instead the turnover mass for which the decrease in $f_{\rm CGM}$ occurs is pushed to higher or lower masses. Other parameters are interesting in that for some values they drastically increase the sharpness of the drop in $f_{\rm CGM}$ such as the high parameter variations of the IMF slop and the Wind Density Factor and the low value of $A_{\rm SN1}$.

We can begin to contextualize these results by comparing to \cite{camels_data_release2} who for the CAMELS-TNG 1P-28 set examined three key quantities: the cosmic star formation rate density (SFRD), the stellar to halo mass relation ($M_*/M_h$), and the gas power spectrum $P_{\rm gas}(k)$ normalized by that of the fiducial model. They find that for SFRD the five most important parameters are: $\Omega_m$, $\sigma_8$, $A_{\rm SN1}$, $h$, $n_s$. For the stellar-to-halo mass relation, they note that all parameters make a visible impact except for $A_{\rm AGN1}$ and $A_{\rm AGN2}$. The three parameters that make the most significant impact are $A_{\rm SN1}$, $A_{\rm SN2}$, and the Wind Free Travel Density Factor. The degree of impact of some of these parameters is in line with our findings.

Focusing on individual parameters that display particularly interesting and sometimes seemingly counterintuitive trends, we provide brief explanations of WindDensFac, IMFSlope, SNIIMinMass, WindERedcFac, BHFeedbackFac, BHRadEff, QuasarThresh, and QuasarThrPower.

Increasing WindDensFac and increasing IMFSlope have similar trends to each other, as well as decreasing $A_{\rm SN1}$, which reduces the mass loading factor and energy content per unit SFR. The increase of WindDensFactor makes winds more difficult to launch; hence it acts similarly to decreasing $A_{\rm SN1}$. Increasing IMFSlope to a higher number (-1.8) produces more massive stars and supernovae, yet this results in a counter-intuitive trend of making winds more difficult to launch. \cite{Lee_2024} found that the increase in metallicity actually increased the cooling efficiency and while the number of SNe increases, their average energy actually decreases in the TNG model \citep{pillepich_illustris_sim_2018}, leading to a reduced effective mass and energy loading.  

Increasing SNIIMinMass reduces the number of SNe and lowers stellar feedback.  The lower $E_{\rm FB}$,total at all masses results in less stellar feedback, resulting in faster SMBH growth and faster onset of the kinetic mode, which is more capable of clearing the CGM.  Interestingly, we do not see a divergence at low-mass like lowering $A_{\rm SN1}$, and (maybe) the driver is the reduction in metal production from increasing SNIIMinMass does not increase the metallicity and lead to runaway cooling like it does for decreased $A_{\rm SN1}$.  

Increasing WindERedcFac reduces the change in the metallicity dependence of the feedback so that higher metallicity SNe are of more comparable efficiency relative to their low metallicity counterparts.  This makes stellar feedback more efficient at high mass, and delays the onset of kinetic AGN feedback clearing halos.  It has less of an effect at lower mass and metallicity.  

Increasing BHFeedbackFac ($\epsilon_{\rm f,high}$) and BHRadEff ($\epsilon_r$) both increase the efficiency of AGN feedback, which leads to greater $f_{\rm CGM}$.  Discussing the former, BHFeedbackFac increases the efficiency of thermal AGN feedback, and this preventatively suppresses further accretion onto the SMBH, hence delaying the onset of the baryon-clearing kinetic mode.  The latter, BHRadEff is actually a similar parameter, but directly suppresses SMBH growth by radiating away that rest mass energy and using the fiducial BHFeedbackFac value.  Hence, BHRadEff has the dual effect of reducing SMBH growth directly and preventatively; therefore the onset of kinetic feedback is even later (earlier) for a higher (lower) value compared to BHFeedbackFac.  

QuasarThresh is the normalization of the threshold of the Eddington ratio to enter kinetic-mode feedback.  A lower (higher) value makes it harder (easier) to enter the low accretion mode, while the high accretion mode, which continues to grow SMBH without clearing CGM, is more (less) prevalent leading to more (less) overall $E_{\rm FB}$.  

Setting QuasarThrPower to its lowest value, $\beta=0$, leads to no mass dependence for BH to reach kinetic mode, so it is far easier for low-mass galaxies to reach kinetic mode and CGM clearing proceeds efficiently at low mass, while it is harder for high-mass galaxies to achieve kinetic mode.  $f_{\rm CGM}$ is very sensitive to low values of this parameter, since it effectively takes away the $10^8 M_{\odot}$ pivot mass for kinetic mode.  QuasarThrPower set to higher values creates even more sensitivity to this pivot mass, which does not change much more the fiducial case, but makes a slightly sharper onset of CGM clearing.

Overall, our analysis of the parameters of the CAMELS-IllustrisTNG 1P-28 set highlights the complex and sometimes counterintuitive ways in which subgrid model feedback parameters can effect the cumulative feedback energetics and halo properties such as the CGM gas fraction. Our results underscore the complex interplay between SNe and AGN feedback mechanisms and the non-linear nature of galaxy formation models. As we refine theoretical models and simulations of galaxy evolution, we must work to better understand these dynamics.

\bibliography{fb_ref}{}
\bibliographystyle{aasjournal}

\end{document}